\input harvmac\skip0=\baselineskip
\input epsf

\newcount\figno
\figno=0
\def\fig#1#2#3{
\par\begingroup\parindent=0pt\leftskip=1cm\rightskip=1cm\parindent=0pt
\baselineskip=11pt \global\advance\figno by 1 \midinsert
\epsfxsize=#3 \centerline{\epsfbox{#2}} \vskip 12pt {\bf Fig.\
\the\figno: } #1\par
\endinsert\endgroup\par
}
\def\figlabel#1{\xdef#1{\the\figno}}
\def\encadremath#1{\vbox{\hrule\hbox{\vrule\kern8pt\vbox{\kern8pt
\hbox{$\displaystyle #1$}\kern8pt} \kern8pt\vrule}\hrule}}



\lref\osv{
  H.~Ooguri, A.~Strominger and C.~Vafa,
  ``Black hole attractors and the topological string,''
  Phys.\ Rev.\ D {\bf 70}, 106007 (2004)
  [arXiv:hep-th/0405146].
}

\lref\MaldacenaRE{
  J.~M.~Maldacena,
  ``The large N limit of superconformal field theories and supergravity,''
  Adv.\ Theor.\ Math.\ Phys.\  {\bf 2}, 231 (1998)
  [Int.\ J.\ Theor.\ Phys.\  {\bf 38}, 1113 (1999)]
  [arXiv:hep-th/9711200].
}

\lref\ItzhakiDD{
  N.~Itzhaki, J.~M.~Maldacena, J.~Sonnenschein and S.~Yankielowicz,
  ``Supergravity and the large N limit of theories with sixteen
  Phys.\ Rev.\  D {\bf 58}, 046004 (1998)
  [arXiv:hep-th/9802042].
}

\lref\KlebanovJA{
  I.~R.~Klebanov and A.~M.~Polyakov,
  ``AdS dual of the critical O(N) vector model,''
  Phys.\ Lett.\  B {\bf 550}, 213 (2002)
  [arXiv:hep-th/0210114].
}

\lref\SchwarzYJ{
  J.~H.~Schwarz,
  ``Superconformal Chern-Simons theories,''
  JHEP {\bf 0411}, 078 (2004)
  [arXiv:hep-th/0411077].
}

\lref\GubserTV{
  S.~S.~Gubser, I.~R.~Klebanov and A.~M.~Polyakov,
  ``A semi-classical limit of the gauge/string correspondence,''
  Nucl.\ Phys.\  B {\bf 636}, 99 (2002)
  [arXiv:hep-th/0204051].
}

\lref\ChenEE{
  W.~Chen, G.~W.~Semenoff and Y.~S.~Wu,
  ``Two loop analysis of nonAbelian Chern-Simons theory,''
  Phys.\ Rev.\  D {\bf 46}, 5521 (1992)
  [arXiv:hep-th/9209005].
}

\lref\GauntlettNG{
  J.~P.~Gauntlett, N.~Kim and D.~Waldram,
  ``M-fivebranes wrapped on supersymmetric cycles,''
  Phys.\ Rev.\  D {\bf 63}, 126001 (2001)
  [arXiv:hep-th/0012195].
}

\lref\IvanovFN{
  E.~A.~Ivanov,
  ``Chern-Simons matter systems with manifest N=2 supersymmetry,''
  Phys.\ Lett.\  B {\bf 268}, 203 (1991).
}

\lref\ZupnikEN{
  B.~M.~Zupnik and D.~G.~Pak,
  ``SUPERFIELD FORMULATION OF THE SIMPLEST THREE-DIMENSIONAL GAUGE THEORIES AND
  CONFORMAL SUPERGRAVITIES,''
  Theor.\ Math.\ Phys.\  {\bf 77}, 1070 (1988)
  [Teor.\ Mat.\ Fiz.\  {\bf 77}, 97 (1988)].
}

\lref\KapustinMT{
  A.~N.~Kapustin and P.~I.~Pronin,
  ``Nonrenormalization theorem for gauge coupling in (2+1)-dimensions,''
  Mod.\ Phys.\ Lett.\  A {\bf 9}, 1925 (1994)
  [arXiv:hep-th/9401053].
}

\lref\KapustinHA{
  A.~Kapustin and M.~J.~Strassler,
  ``On mirror symmetry in three dimensional Abelian gauge theories,''
  JHEP {\bf 9904}, 021 (1999)
  [arXiv:hep-th/9902033].
}

\lref\KaoGF{
  H.~C.~Kao, K.~M.~Lee and T.~Lee,
  ``The Chern-Simons coefficient in supersymmetric Yang-Mills Chern-Simons
  theories,''
  Phys.\ Lett.\  B {\bf 373}, 94 (1996)
  [arXiv:hep-th/9506170].
}

\lref\McGreevyHK{
  J.~McGreevy, E.~Silverstein and D.~Starr,
  ``New dimensions for wound strings: The modular transformation of geometry to
  topology,''
  Phys.\ Rev.\  D {\bf 75}, 044025 (2007)
  [arXiv:hep-th/0612121].
}

\lref\AharonySX{
  O.~Aharony, J.~Marsano, S.~Minwalla, K.~Papadodimas and M.~Van Raamsdonk,
  ``The Hagedorn / deconfinement phase transition in weakly coupled large N
  gauge theories,''
  Adv.\ Theor.\ Math.\ Phys.\  {\bf 8}, 603 (2004)
  [arXiv:hep-th/0310285].
}

\lref\AlvarezGaumeWK{
  L.~Alvarez-Gaume, J.~M.~F.~Labastida and A.~V.~Ramallo,
  ``A Note on Perturbative Chern-Simons Theory,''
  Nucl.\ Phys.\  B {\bf 334}, 103 (1990).
}

\lref\SchnitzerQT{
  H.~J.~Schnitzer,
  ``Confinement / deconfinement transition of large N gauge theories with  N(f)
  fundamentals: N(f)/N finite,''
  Nucl.\ Phys.\  B {\bf 695}, 267 (2004)
  [arXiv:hep-th/0402219].
}

\lref\KorchemskyWG{
  G.~P.~Korchemsky and A.~V.~Radyushkin,
  ``Renormalization of the Wilson Loops Beyond the Leading Order,''
  Nucl.\ Phys.\  B {\bf 283}, 342 (1987).
}

\lref\KorchemskyXV{
  G.~P.~Korchemsky and G.~Marchesini,
  ``Structure function for large x and renormalization of Wilson loop,''
  Nucl.\ Phys.\  B {\bf 406}, 225 (1993)
  [arXiv:hep-ph/9210281].
}

\lref\AvdeevZA{
  L.~V.~Avdeev, G.~V.~Grigorev and D.~I.~Kazakov,
  ``Renormalizations in Abelian Chern-Simons field theories with matter,''
  Nucl.\ Phys.\  B {\bf 382}, 561 (1992).
}

\lref\AvdeevJT{
  L.~V.~Avdeev, D.~I.~Kazakov and I.~N.~Kondrashuk,
  ``Renormalizations in supersymmetric and nonsupersymmetric nonAbelian
  Chern-Simons field theories with matter,''
  Nucl.\ Phys.\  B {\bf 391}, 333 (1993).
}

\lref\KaoIG{
  H.~C.~Kao and K.~M.~Lee,
  ``Selfdual Chern-Simons systems with an N=3 extended supersymmetry,''
  Phys.\ Rev.\  D {\bf 46}, 4691 (1992)
  [arXiv:hep-th/9205115].
}

\lref\KaoGS{
  H.~C.~Kao,
  ``Selfdual Yang-Mills Chern-Simons Higgs systems with an N=3 extended
  supersymmetry,''
  Phys.\ Rev.\  D {\bf 50}, 2881 (1994).
}

\lref\DrukkerRR{
  N.~Drukker and D.~J.~Gross,
  ``An exact prediction of N = 4 SUSYM theory for string theory,''
  J.\ Math.\ Phys.\  {\bf 42}, 2896 (2001)
  [arXiv:hep-th/0010274].
}

\lref\WittenHF{
  E.~Witten,
  ``Quantum field theory and the Jones polynomial,''
  Commun.\ Math.\ Phys.\  {\bf 121}, 351 (1989).
}

\lref\vandeWeteringXP{
  J.~F.~W.~van de Wetering,
  ``Knot invariants and universal R matrices from perturbative Chern-Simons
  theory in the almost axial gauge,''
  Nucl.\ Phys.\  B {\bf 379}, 172 (1992).
}

\lref\CorradoNV{
  R.~Corrado, K.~Pilch and N.~P.~Warner,
  Nucl.\ Phys.\  B {\bf 629}, 74 (2002)
  [arXiv:hep-th/0107220].
}

\lref\AhnZY{
  C.~h.~Ahn and S.~J.~Rey,
  ``More CFTs and RG flows from deforming M2/M5-brane horizon,''
  Nucl.\ Phys.\  B {\bf 572}, 188 (2000)
  [arXiv:hep-th/9911199].
}

\lref\AhnAQ{
  C.~h.~Ahn and J.~Paeng,
  ``Three-dimensional SCFTs, supersymmetric domain wall and renormalization
  group flow,''
  Nucl.\ Phys.\  B {\bf 595}, 119 (2001)
  [arXiv:hep-th/0008065].
}

\lref\AharonyTI{
  O.~Aharony, S.~S.~Gubser, J.~M.~Maldacena, H.~Ooguri and Y.~Oz,
  ``Large N field theories, string theory and gravity,''
  Phys.\ Rept.\  {\bf 323}, 183 (2000)
  [arXiv:hep-th/9905111].
}

\lref\GubserBC{
  S.~S.~Gubser, I.~R.~Klebanov and A.~M.~Polyakov,
  ``Gauge theory correlators from non-critical string theory,''
  Phys.\ Lett.\  B {\bf 428}, 105 (1998)
  [arXiv:hep-th/9802109].
}

\lref\WittenQJ{
  E.~Witten,
  ``Anti-de Sitter space and holography,''
  Adv.\ Theor.\ Math.\ Phys.\  {\bf 2}, 253 (1998)
  [arXiv:hep-th/9802150].
}

\lref\BeisertIB{
  N.~Beisert, R.~Hernandez and E.~Lopez,
  ``A crossing-symmetric phase for AdS(5) x S**5 strings,''
  JHEP {\bf 0611}, 070 (2006)
  [arXiv:hep-th/0609044].
}

\lref\BeisertEZ{
  N.~Beisert, B.~Eden and M.~Staudacher,
  ``Transcendentality and crossing,''
  J.\ Stat.\ Mech.\  {\bf 0701}, P021 (2007)
  [arXiv:hep-th/0610251].
}

\lref\EdenRX{
  B.~Eden and M.~Staudacher,
  ``Integrability and transcendentality,''
  J.\ Stat.\ Mech.\  {\bf 0611}, P014 (2006)
  [arXiv:hep-th/0603157].
}

\lref\HofmanXT{
  D.~M.~Hofman and J.~M.~Maldacena,
  ``Giant magnons,''
  J.\ Phys.\ A  {\bf 39}, 13095 (2006)
  [arXiv:hep-th/0604135].
}

\lref\MaldacenaRV{
  J.~Maldacena and I.~Swanson,
  ``Connecting giant magnons to the pp-wave: An interpolating limit of AdS(5) x
  S**5,''
  arXiv:hep-th/0612079.
}

\lref\DoreyXN{
  N.~Dorey, D.~M.~Hofman and J.~Maldacena,
  ``On the singularities of the magnon S-matrix,''
  arXiv:hep-th/0703104.
}

\lref\AldayQF{
  L.~F.~Alday, G.~Arutyunov, M.~K.~Benna, B.~Eden and I.~R.~Klebanov,
  ``On the strong coupling scaling dimension of high spin operators,''
  arXiv:hep-th/0702028.
}

\lref\FrolovQE{
  S.~Frolov, A.~Tirziu and A.~A.~Tseytlin,
  ``Logarithmic corrections to higher twist scaling at strong coupling from
  AdS/CFT,''
  Nucl.\ Phys.\  B {\bf 766}, 232 (2007)
  [arXiv:hep-th/0611269].
}

\lref\MinahanBD{
  J.~A.~Minahan, A.~Tirziu and A.~A.~Tseytlin,
  ``Infinite spin limit of semiclassical string states,''
  JHEP {\bf 0608}, 049 (2006)
  [arXiv:hep-th/0606145].
}

\lref\BerensteinJQ{
  D.~Berenstein, J.~M.~Maldacena and H.~Nastase,
  ``Strings in flat space and pp waves from N = 4 super Yang Mills,''
  JHEP {\bf 0204}, 013 (2002)
  [arXiv:hep-th/0202021].
}

\lref\BeisertRY{
  N.~Beisert,
  ``The dilatation operator of N = 4 super Yang-Mills theory and
  integrability,''
  Phys.\ Rept.\  {\bf 405}, 1 (2005)
  [arXiv:hep-th/0407277].
}

\lref\BeisertYS{
  N.~Beisert,
  ``The su(2|3) dynamic spin chain,''
  Nucl.\ Phys.\  B {\bf 682}, 487 (2004)
  [arXiv:hep-th/0310252].
}

\lref\AlvarezGaumeIG{
  L.~Alvarez-Gaume and E.~Witten,
  ``Gravitational Anomalies,''
  Nucl.\ Phys.\  B {\bf 234}, 269 (1984).
}

\lref\DouglasES{
  M.~R.~Douglas and S.~Kachru,
  ``Flux compactification,''
  arXiv:hep-th/0610102.
}

\lref\RedlichDV{
  A.~N.~Redlich,
  ``Parity Violation And Gauge Noninvariance Of The Effective Gauge Field
  Action In Three-Dimensions,''
  Phys.\ Rev.\  D {\bf 29}, 2366 (1984).
}

\lref\WittenDS{
  E.~Witten,
  ``Supersymmetric index of three-dimensional gauge theory,''
  arXiv:hep-th/9903005.
}

\lref\EllisTY{
  R.~K.~Ellis, H.~Georgi, M.~Machacek, H.~D.~Politzer and G.~G.~Ross,
  ``Perturbation Theory And The Parton Model In QCD,''
  Nucl.\ Phys.\  B {\bf 152}, 285 (1979).
}

\lref\BelitskyYS{
  A.~V.~Belitsky, A.~S.~Gorsky and G.~P.~Korchemsky,
  ``Gauge / string duality for QCD conformal operators,''
  Nucl.\ Phys.\  B {\bf 667}, 3 (2003)
  [arXiv:hep-th/0304028].
}

\lref\KruczenskiFB{
  M.~Kruczenski,
  ``A note on twist two operators in N = 4 SYM and Wilson loops in Minkowski
  signature,''
  JHEP {\bf 0212}, 024 (2002)
  [arXiv:hep-th/0210115].
}

\noblackbox

\Title{\vbox{\baselineskip12pt\hbox{} }} { Notes on Superconformal
Chern-Simons-Matter Theories} \centerline{Davide Gaiotto\foot{Email:
dgaiotto@gmail.com} and Xi Yin\foot{Email: xiyin@fas.harvard.edu} }
\smallskip
\centerline{Jefferson Physical Laboratory, Harvard University,
Cambridge, MA 02138} \vskip .6in \centerline{\bf Abstract} { The
three dimensional ${\cal N}=2$ supersymmetric Chern-Simons theory
coupled to matter fields, possibly deformed by a superpotential,
give rise to a large class of exactly conformal theories with
Lagrangian descriptions. These theories can be arbitrarily weakly
coupled, and hence can be studied perturbatively. We study the
theories in the large $N$ limit, and compute the two-loop anomalous
dimension of certain long operators. Our result suggests that
various ${\cal N}=2$ $U(N)$ Chern-Simons theories coupled to
suitable matter fields are dual to open or closed string theories in
$AdS_4$, which are not yet constructed. } \vskip .3in

\Date{April 2007}

\listtoc \writetoc

\newsec{Introduction}

The $AdS/CFT$ correspondence
\refs{\MaldacenaRE,\GubserBC,\WittenQJ,\AharonyTI} has led to
tremendous insight into gauge theory as well as non-perturbative
descriptions of theories of gravity. Much progress has been made in
understanding $AdS_5/CFT_4$ and $AdS_3/CFT_2$, where both the string
theory side and the conformal field theory side can be solved in
certain limits. (For a sample of recent advances see
\refs{\BeisertIB, \BeisertEZ, \EdenRX, \HofmanXT, \MaldacenaRV,
\DoreyXN, \AldayQF, \FrolovQE, \MinahanBD}.) The $AdS_4/CFT_3$
correspondence, on the other hand, is much less understood. Two
primary classes of examples of $AdS_4/CFT_3$ have been studied so
far: (1) the duality between M-theory on $AdS_4\times S^7$ (as well
as $AdS_4\times X_7$ with other Sasaki-Einstein seven-manifolds
$X_7$) and the M2-brane CFT, which is believed to be the infrared
fixed point of the three dimensional ${\cal N}=8$ $U(N)$
super-Yang-Mills theory \ItzhakiDD; (2) the duality between higher
spin gauge theory in $AdS_4$ and the IR fixed point of three
dimensional $O(N)$ model \KlebanovJA. In the first example, it is
difficult to calculate anything in the IR CFT from the gauge theory,
which is strongly coupled. For instance, one does not even
understand from the gauge theory perspective why the IR CFT has
$N^{3/2}$ degrees of freedom, as predicted from the gravity dual.
The gravity side, which involves M-theory, is also not completely
formulated. In general we do not know how to go beyond supergravity
and semi-classical M-branes. In the second example, one can do
computations in the CFT in $1/N$ expansion and $\epsilon$-expansion,
but the gravity dual is again difficult to analyze. In both classes
of examples, there is no adjustable parameter in the theory other
than $N$ itself.

It was pointed out in \SchwarzYJ\ that supersymmetric Chern-Simons
theories, which are by themselves topological but may be coupled to
matter fields that carry physical degrees of freedom, give rise to a
natural class of classically conformal theories. Furthermore, since
the Chern-Simons level is not renormalized up to a possible 1-loop
shift, such theories are anticipated to be exactly conformal quantum
mechanically, and hence potentially giving rise to an interesting
new class of $AdS_4/CFT_3$ correspondences. The non-renormalization properties
of ${\cal N}=2$ and ${\cal N}=3$ Chern-Simons theories coupled to matter fields
have been previously studied in \refs{\KapustinMT, \ChenEE,\AvdeevZA, \AvdeevJT,\KapustinHA}.

More precisely, ${\cal N}=2$ and ${\cal N}=3$
supersymmetric Chern-Simons theories coupled minimally to matter
fields are exactly conformal, in the sense that there is no relevant
or marginal quantum corrections to the classical action. The
theories are labeled by the gauge group $G$, matter representation
$R$, and the Chern-Simons level $k$. $1/k$ plays the role of the
coupling constant, and in particular the theory can be made
arbitrarily weakly coupled and can be analyzed in conventional
perturbation theory. As we will show, the ${\cal N}=3$ theory can be obtained
as the IR fixed point of the ${\cal N}=2$ theory perturbed by a certain
superpotential. In fact, there is an even larger class of
superconformal field theories, obtained from more general
superpotential deformations of the ${\cal N}=2$ CS-matter theory,
all of which have (weakly coupled) Lagrangian descriptions.

The supersymmetric Chern-Simons theories coupled to fundamental or
adjoint matter fields can often be engineered as the IR limit of the
world volume theory on branes in string theory. One might try to
find the gravity dual of CS-matter theories by studying the
decoupling limit of such brane solutions. However, we do not know
any example in which such decoupling limit exists as a smooth
$AdS_4$ solution in supergravity. It is conceivable that the $AdS_4$
dual of CS-matter theories, if described by a weakly coupled string
theory, always has radius at string scale.

While we do not know a direct construction of the gravity dual, we
can still study the large $N$ limit of CS-matter theories
perturbatively, and look for signatures of strings. It is a priori
not obvious whether the usual lore of gauge/string duality applies
in this case, since the Chern-Simons gauge field is effectively
infinitely massive and does not carry propagating degrees of freedom
by itself. Take as an example the ${\cal N}=2$ $U(N)$ Chern-Simons
theory coupled to $N_f$ fundamental matter fields. We expect this
CFT to be dual to a theory of gravity in $AdS_4$, with $U(N_f)$
gauge fields. One can consider twist-1 operators of the form
\eqn\twisttt{ \bar\phi D_{(\mu_1}\cdots D_{\mu_n)}\phi } at large
spin $n$, where $\phi$ is the scalar matter field and $D_\mu$ are
gauge covariant derivatives. At weak 't Hooft coupling $\lambda\sim
N/k$, the leading contribution to the anomalous comes in at
two-loop. To this order we find that there are corrections of order
${\cal O}(\lambda^2)$, bounded in the $n\to \infty$ limit, and
corrections of order ${\cal O}(\lambda^2 N_f/N)$, which grow like
$\ln(n)$ at large $n$. The latter is the expected growth of the
anomalous dimension of the operator dual to a classical string
spinning in $AdS_4$. This sort of agreement was found in four
dimensional ${\cal N}=4$ super-Yang-Mills \GubserTV, where the
$\ln(n)$ growth in fact holds to all orders in the 't Hooft
coupling. This suggests that in the large $N$ limit, with $N_f/N$
finite, \twisttt\ could indeed be dual to a classical spinning open
string in $AdS_4$, and that ${\cal N}=2$ $U(N)$ Chern-Simons theory
with fundamental matter could be dual to a open string theory in
$AdS_4$ (i.e. with $N_f$ space filling D-branes). The radius of
$AdS_4$ in string units will be a function of $\lambda$ and $N_f/N$.

The above considerations extend naturally to theories with adjoint
matter. It is natural to study spin chain descriptions of long
operators in these theories. For example, the ${\cal N}=3$ theory
with one adjoint matter has $SU(2)_R\times SU(2)_f$ global symmetry.
There is a corresponding $SU(2)_R\times SU(2)_f$ spin chain, which
turns out to be non-integrable at two-loop. In general we do not
expect the spin chain associated with CS-matter theories to be
integrable. Although we do not know the precise holographic dual of
this theory, our findings are compatible with a 7-dimensional
supergravity dual at large 't Hooft coupling, with the geometry of
the coset $OSp(3|4)/SO(3,1)$. Indeed, the spectrum of protected
operators consists of a tower of irreducible representations of
$OSp(3|4) \times SU(2)_f$ with spin $(j,j)$ under $SU(2)_R\times
SU(2)_f$ for each $j$, as expected for the KK-tower of modes on
$S^3$. The analysis of the spin chain reinforces this idea: giant
magnon-like excitations of large R-charge which carry spin
$(j-\half,j)$ have the same anomalous dimension as excitations of
spin $(j,j-\half)$. This is the behavior expected for a giant magnon
moving along the equator of the $S^3$.

 This paper is organized as follows. In section 2 we
recall the ${\cal N}=2$ and ${\cal N}=3$ Chern-Simons-matter
theories, and the arguments for their non-renormalization
properties. We will show that the ${\cal N}=2$ theories deformed by
superpotentials can flow to other weakly coupled superconformal
fixed points, including ${\cal N}=3$ ones. In section 3 we study the
't Hooft limit of various CS-matter theories. For the abelian
theory, we will solve the free energy at large $N_f$ and study the
thermodynamics of the theory. We will then discuss the operator
spectrum of the $U(N)$ Chern-Simons theory coupled to fundamental
matter. In particular, we compute the two-loop anomalous dimension
of twist-1 operators at large spin. We will also discuss
supersymmetric Wilson loops and spin chain descriptions of long
operators. Section 4 contains some comments on the possible string
theory $AdS_4$ dual. Appendix A summarizes some details of the
${\cal N}=3$ Lagrangian. Appendix B presents the saddle point
analysis of the abelian theory in the large $N_f$ limit. Appendix C
contains a discussion of the free nonabelian theory on the sphere.
In appendix D, we present the computation of the two-loop anomalous
dimensions of some operators.

\newsec{Supersymmetric Chern-Simons theories coupled to matter fields}

\subsec{${\cal N}=2$ Chern-Simons-matter theory}

In this subsection we review the ${\cal N}=2$ Chern-Simons theory
coupled to matter fields
\refs{\ZupnikEN,\IvanovFN,\AvdeevZA,\AvdeevJT}, as well as its
non-renormalization properties. We shall start by describing the
Lagrangian in superspace. The three dimensional ${\cal N}=2$ vector
superfield $V$ consists of the gauge field $A_\mu$, an auxiliary
scalar field $\sigma$, a two-component Dirac spinor $\chi$, and
another scalar $D$. The superspace Lagrangian for abelian ${\cal
N}=2$ Chern-Simons theory, coupled to matter chiral superfield
$\Phi$, is given by \eqn\abcsa{ S^{{\cal N}=2}_{Ab}=\int d^3x \int
d^4\theta \left({k\over 4\pi} V \Sigma +\bar\Phi e^V \Phi \right)  }
where $\Sigma=\bar D^\alpha D_\alpha V$. When there are several
matter fields $\Phi_i$ of different charges, it is convenient to
absorb $k$ into the charge and write the action as \eqn\abcsamore{
S^{{\cal N}=2}_{Ab}=\int d^3x \int d^4\theta \left( V \Sigma
+\bar\Phi_i e^{q_iV} \Phi_i \right)  } The nonabelian ${\cal N}=2$
Chern-Simons action is trickier to write in terms of the nonabelian
vector superfield $V$: \eqn\suppac{ S^{{\cal N}=2} = \int d^3x \int
d^4\theta \left\{ {k\over 2\pi}\int_0^1 dt{\rm Tr} \left[V \bar
D^\alpha (e^{-tV} D_\alpha e^{tV})\right]+\bar\Phi e^V \Phi \right\}
} where the matter field $\Phi$ is in an arbitrary representation
$R$ of the gauge group. Here ``${\rm Tr}$" is normalized to be the trace in
the fundamental representation when the gauge group is
$U(N)$ or $SU(N)$. We will denote by ${\rm tr}_R$ the trace taken in
representation $R$. In component fields (Wess-Zumino
gauge), the Chern-Simons action is simply
 \eqn\asdffcd{ S_{CS}^{{\cal N}=2} = {k\over 4\pi}\int {\rm Tr} (A\wedge dA+{2\over
3}A^3-\bar \chi \chi+2D\sigma)} The level $k$ is quantized to be
integer valued\foot{or half-integer valued when there is a ``parity
anomaly" \refs{\AlvarezGaumeIG,\RedlichDV}, depending on the matter
content. This subtlety is not essential for us as we will mostly
work with weak coupling $k\gg 1$ and/or 't Hooft-like limits.} to
ensure invariance under large gauge transformations. We will often
consider $N_f$ ${\cal N}=2$ chiral matter fields in some
representation $R$, denoted by $\Phi^i=(\phi^i,\psi^i)$, with global
$U(N_f)$ flavor symmetry. The coupling to the vector multiplet is
the standard one, \eqn\matte{ \eqalign{ & S_{matter}=\int d^4\theta
\sum_{i=1}^{N_f}\bar\Phi^i e^V \Phi^i = \int \sum_{i=1}^{N_f}
\left(D_\mu \bar \phi^i D^\mu \phi^i +i\bar\psi^i \gamma^\mu D_\mu
\psi^i -\bar \phi^i\sigma^2 \phi^i+\bar \phi^i D \phi^i\right. \cr
&~~~\left.-\bar\psi^i\sigma\psi^i+i\bar \phi^i \bar \chi\psi^i -i
\bar\psi^i \chi  \phi^i \right). } } The auxiliary fields $\sigma$ and
$D$ are understood to act on $(\phi,\psi)$ in the representation
$R$. Integrating out $D$ sets $\sigma = -{4\pi\over k} (\bar \phi^i T^a \phi^i) t^a$,
where $T^a$ are generators of the Lie algebra of the gauge group,
normalized so that ${\rm Tr}(T^a T^b)=\half \delta^{ab}$. Further integrating out
$\chi,\bar\chi$ yields the action \eqn\adct{ \eqalign{ & S^{{\cal
N}=2}={k\over 4\pi} CS(A)+\int D_\mu \bar \phi^i D^\mu \phi^i
-{16\pi^2\over k^2} (\bar \phi^i T^a \phi^i) (\bar \phi^j
T^b \phi^j) (\bar \phi^k T^a T^b \phi^k) \cr & ~~+
i\bar\psi^i \gamma^\mu D_\mu \psi^i -{4\pi\over k} (\bar \phi^i T^a \phi^i) (\bar \psi^j T^a \psi^j) -
{8\pi\over k} (\bar\psi^i T^a \phi^i) (\bar \phi^j T^a
\psi^j). } } This action is clearly classically marginal. We will
now argue that it is in fact quantum mechanically exactly marginal,
in the sense that there is no relevant or marginal quantum
corrections to the action that cannot be absorbed into a
redefinition of the fields.

When the matter field $\Phi$ transforms in a irreducible
representation $R$, there is a $U(1)$ symmetry $\Phi\to
e^{i\alpha}\Phi$. Since there is no anomaly for continuous global
symmetry in three dimensions, this $U(1)$ symmetry holds in the
quantum theory and forbids any superpotential involving $\Phi$ by
holomorphy. Similarly, when the matter field lies in a reducible
representation of the gauge group, there are $U(1)$ symmetries
acting on each irreducible part of $\Phi$, and the same argument
forbids a dynamically generated superpotential.

It is well known that the Chern-Simons level $k$ is not renormalized
beyond a possible finite 1-loop shift \KapustinMT. \foot{See
\ChenEE\ for a discussion on the regularization dependence of the
1-loop shift.} The simplest way to argue this is that $k$ is
quantized to be integer valued in order for the path integral to be
invariant under large gauge transformations \WittenDS. Any quantum
correction to $k$ at 2-loop or higher order will be suppressed by
$1/k$, which in general cannot be integer valued.

So the only possible quantum corrections to the classical Lagrangian
is to the Kahler potential. This indeed happens. However, any
corrections to the K\"ahler potential are either irrelevant in the
IR or can be absorbed by a rescaling of $\Phi$.\foot{In Wilsonian
effective action all corrections to the K\"ahler potential are
non-singular at $\Phi=0$, hence this argument is valid. This would
not be the case for the 1PI effective action, since we would be
integrating out massless fields, and the effective K\"ahler
potential may well be singular at $\Phi=0$.} One might also worry
about possible generation of Fayet-Iliopoulos term, when there are
matter fields charged under the $U(1)$ part of the gauge field. Any
dynamical FI parameter must be of the form $D_\alpha\bar
D^\alpha(\cdots)$ in order to preserve gauge invariance, where
$(\cdots)$ is a gauge invariant combination of the fields. Again,
such terms are irrelevant in the IR. In conclusion, there cannot be
any relevant or marginal correction to the classical Lagrangian
\adct, and hence the theory is exactly marginal.

The vanishing of two-loop beta function of the matter couplings has
been explicitly shown in \AvdeevZA\ for the abelian theory and
\AvdeevJT\ for the nonabelian theory.


As a consistency check of the existence of these conformal field
theories, let us consider the example of a $U(1)$ CS-matter theory
with both positively and negatively charged matter fields (say of
charges $q$ and $-q$). The scalar potential takes the form
\eqn\scpotaa{ V(\phi,\tilde\phi)={q^4\over
4}(|\phi|^2-|\tilde\phi|^2)^2(|\phi|^2+|\tilde\phi|^2) } A generic
point on the Higgs branch moduli space is parameterized by nonzero
$\phi$. This moduli space cannot be lifted since no superpotential
can be generated. Writing $\tilde\phi=\phi e^{\rho+i\theta}$, $\rho$
has mass of order $q^2|\phi|^2$ and $\theta$ is a Goldstone boson
which can absorbed by a gauge transformation. Similarly one fermion
gets massive and the other remains massless. One obtains the quantum
corrected metric on the moduli space by integrating out the massive
fields. For example, the leading (two-loop) correction to the
kinetic term for $\phi$ takes the form $q^4|\partial\phi|^2
\ln(|\phi|^2/\mu)$. Such corrections result in a cone-shaped metric
near $\phi=0$, reflecting the anomalous dimension of $\phi$. The
distance from the origin $\phi=0$ to a generic point on the Higgs
branch moduli space is finite, at least for small $q$. If a term
like ${\mu\over|\phi|^2}|\partial\phi|^2$ were generated, it would
suggest that the CFT at $\phi=0$ may not exist, but this doesn't
seem to happen in the CS-matter theory, at least at weak coupling.

There may be caveats in our argument for the theory at strong
coupling. In the ${\cal N}=2$ theory, generally, the $U(1)_R$ charge
of $\phi$ gets renormalized, as we will discuss later. At strong
coupling there could be dangerously irrelevant terms in the K\"ahler
potential being generated, spoiling the above non-renormalization
argument. We also see such possibility from the above analysis of
the Higgs branch moduli space (when exists): if the wave function
renormalization of $\phi$ is sufficiently large, the metric on the
moduli space could be such that the origin is at infinite distance,
and the CFT may cease to exist. It is not clear to us if this
actually happens. The ${\cal N}=3$ theory, which we will describe in
the next subsection, does not have R-charge renormalization nor wave
function renormalization. So one might expect better behavior of the
theory at strong coupling.

\subsec{${\cal N}=3$ Chern-Simons-matter theory}

The maximally supersymmetric extension of Chern-Simons theory
appears to be ${\cal N}=3$ \refs{\KaoGF,\KapustinHA}, which can be
coupled to hypermultiplet matter fields. In ${\cal N}=2$ language
the matter fields consists of a pair of chiral multiplets $(Q,\tilde
Q)$, transforming in conjugate representations of the gauge group.
The action for the ${\cal N}=3$ Chern-Simons matter theory takes the
form\foot{Our action may appear different from the nonabelian ${\cal
N}=3$ action of \KaoGS, but they are in fact the same. For example,
the Chern-Simons part of the ${\cal N}=3$ action in \KaoGS\ contains
the extra term ${\rm Tr}\bar\phi [\sigma,\phi]$. This term can be
absorbed into the scalar potential of \KaoGS, and one recovers our
F-term after redefining $\phi_1$ and $\bar\phi_2$ in \KaoGS\ as $Q$
and $\tilde Q$, respectively. } \eqn\nteac{ S^{{\cal N}=2} =
S_{CS}^{{\cal N}=2} + \int d^4\theta (\overline Q e^V Q+\tilde Q
e^{-V} \overline{\tilde Q}) + \left[\int d^2\theta \left(-{k\over
4\pi}{\rm Tr}\Phi^2 + \tilde Q \Phi Q\right)+c.c.\right] } Here $\Phi$ is an
auxiliary chiral superfield in the adjoint representation, combined
with $V$ to give the ${\cal N}=4$ vector multiplet. The scalar
component of $\Phi$, which we denote by $\phi$, combines with
$\sigma$ to form a triplet under the $SU(2)_R$ symmetry. Similarly,
$F_\Phi$ combines with the auxiliary field $D$ to form a triplet.

We will often consider theories with $N_f$ matter hypermultiplets in
some irreducible representation $R$. There is $U(N_f)$ global flavor
symmetry. When $R$ is a real representation (say the adjoint), the
flavor symmetry is enhanced to $USp(2N_f)$. In appendix A we write
the Lagrangian in a manifestly $SU(2)_R\times USp(2N_f)$ invariant
form.

The auxiliary field $\Phi$ may be simply integrated out, resulting
in a superpotential \eqn\wsurw{ W = {2\pi\over k} (\tilde Q T^a
Q)(\tilde Q T^a Q). }
In other words, the ${\cal N}=3$ Chern-Simons-matter theory is the
same as ${\cal N}=2$ Chern-Simons-matter theory with matter fields
$Q,\tilde Q$ and the superpotential \wsurw. Note that the
coefficient of \wsurw\ is fixed by ${\cal N}=3$ supersymmetry.
Simialr non-renormalization argument as in the previous subsection
also applies to the ${\cal N}=3$ theory. The ${\cal N}=3$ theory in
fact has stronger non-renormalization property, since the $SU(2)_R$
charge of the fields cannot be renormalized, in contrast to the
${\cal N}=2$ theory. For example, the gauge invariant meson operator
$\tilde Q Q$ is a chiral primary, whose dimension is protected. It
follows that there is no wave function renormalization for the
matter fields \KapustinHA. In conclusion, the ${\cal N}=3$
Chern-Simons-matter theory is also exactly conformal.

From the point of view of the ${\cal N}=2$ theory, the appearance of
a conformal fixed point at a finite deformation $W$ is somewhat
unexpected. Let us consider a more general superpotential,
\eqn\stumore{ W = {\alpha\over 2}(\tilde Q T^a Q) (\tilde Q T^a
Q), } with $\alpha$ a non-negative real constant (one can always
absorb the phase of $\alpha$ into a redefinition of the $Q$'s). No
other superpotential terms can be generated by standard arguments
based on holomorphy and $U(1)_R$ symmetry.\foot{We should however be
cautious with the standard non-renormalization arguments involving
the gauge coupling, since in Chern-Simons theory the coupling $k$
cannot be promoted to a dynamical field.} When $\alpha\gg 4\pi/k$
and $k\gg 1$, $W$ dominates the interaction, and the theory in this
limit is essentially the three dimensional Wess-Zumino model with
superpotential $W$. This theory has a positive beta function, i.e.
$\alpha$ decreases going to the IR. The leading two-loop RGE for
$\alpha$ takes the form \eqn\taaa{ \mu{d\alpha^2\over d\mu} =
{b_0\over 16\pi^2} \alpha^2\left[\alpha^2-\left({4\pi\over
k}\right)^2 \right],~~~~~~b_0>0. } The coeffient $b_0$ can be determined from the beta function
of the corresponding WZ model, $b_0={2\over\dim R}\left[({\rm tr}_R T^aT^b)^2+{\rm tr}_R
(T^aT^bT^aT^b)\right]$. This is because the theory has
two conformal fixed points, at $\alpha=0$ and $\alpha=4\pi/k$, and
it is not hard to see that the two-loop correction to $\alpha^2$ is
a quadratic function in $\alpha^2$. Hence we learn that a small
perturbation of the ${\cal N}=2$ theory ($\alpha=0$) by the
superpotential \stumore\ flows to the ${\cal N}=3$ conformal fixed
point ($\alpha={4\pi\over k}$). Knowing that $\alpha=0,{4\pi\over
k}$ are exact fixed points, the 2-loop result \taaa\ suggests that
this RG flow holds in the full theory.

\subsec{Chiral operators and chiral primaries}

Let us first discuss the chiral operators in the {\sl abelian}
${\cal N}=2$ theory. The chiral operators are given by gauge
invariant polynomials in $\phi$'s.\foot{Since $\Sigma=\bar DDV$ is
already gauge invariant, the chiral operator $\bar D_\alpha \Sigma$
is a descendant and will be ignored. This is in contrast with four
dimensional gauge theories, in which $\Sigma$ doesn't exist.} Since
there is no superpotential, there is no relations in the chiral
ring, these chiral operators are also chiral primaries. In theories
where all matter fields have charges of the same sign, there are no
chiral primaries. Let us consider the $N=2$ theory with oppositely
charged matter fields $Q^i$ and $\tilde Q^i$, $i=1,\cdots,N_f$.
Consider a chiral primary ${\cal O}=f(Q,\tilde Q)$, where $f$ is a
polynomial of homogeneous degree $n$. The conformal dimension of
${\cal O}$ is given by the unitarity bound, $\Delta = n q_R$, where
$q_R$ is the $U(1)_R$ charge of $Q$ and $\tilde Q$. Classically
$q_R=\half$, but quantum mechanically it is modified to be less than
$\half$. In fact, we have learned from \taaa\ that the
superpotential perturbation $(\tilde Q Q)^2$ is relevant, and the
$U(1)_R$ charge is renormalized at two-loop to \eqn\rcha{
q_R={1\over 2}-{b_0\over 8k^2}+ {\cal O}({1\over k^4}). } This may
be confirmed by directly computing the anomalous dimension of
$\tilde Q Q$. It is also easy to check at two-loop that the
anomalous dimension of $(\tilde QQ)^n$ is $n$ times that of $\tilde
Q Q$, say to order $N_f/k^2$ (the $\sim n^2$ contributions cancel).

In a general nonabelian ${\cal N}=2$ CS-matter theory with gauge
group $G$ and matter fields $\Phi_i$ in irreducible representation
$R_i$, the chiral primaries are gauge invariant polynomials in the
$\Phi_i$'s. Let us consider a few special cases:

\noindent (1) $G=U(N)$, $R_i={\bf N}$, $i=1,\cdots,N_f$. There are
no chiral primaries (besides the identity operator) in this theory.

\noindent (2) $G=SU(N)$, $R_i={\bf N}$, $i=1,\cdots,N_f$. The chiral
primaries only exist for $N_f\geq N$. They are (generated by) baryon
operators, of the form \eqn\bartoo{ B_{i_1\cdots i_{N_f-N}} =
\epsilon_{i_1\cdots i_{N_f-N} j_1\cdots j_N} \epsilon^{a_1\cdots
a_N} \phi^{j_1}_{a_1}\cdots \phi^{j_N}_{a_N} }

\noindent (3) $G=U(N)$, $R_i={\bf N}$, $\tilde R_i={\bf
\overline{N}}$. The chiral primaries are mesons, $M^i_j=\phi^i_a
\tilde \phi^a_j$.

\noindent (4) $G=SU(N)$, $R_i={\bf adj}$, $i=1,\cdots,N_f$. The
chiral primaries are generated by ``words", i.e. the trace of a
string of $\Phi_i$'s in a particular order up to cyclic permutation,
${\cal O}_{i_1\cdots i_n}={\rm Tr}(\Phi_{i_1}\cdots \Phi_{i_n})$.
Since there is no superpotential, the indices are not necessarily
symmetrized. In the infinite $N$ limit, all the traces are
independent operators, and the number of chiral primaries of
dimension $n$ growths exponentially $\sim (N_f)^n$ for $N_f>1$
(modulo the correction due to cyclic permutations). This is a
peculiar feature. It is curious whether such chiral primaries could
be dual to winding string-like objects in negatively curved spaces,
a la \McGreevyHK.

Let us turn to the ${\cal N}=3$ theory. In this case, there cannot
be any quantum correction to $q_R$, since it is the $U(1)$ part of
the $SU(2)$ R-symmetry. Hence there is no anomalous dimension for
the mesons $\tilde QQ$. However, since there is the superpotential
\wsurw, there are nontrivial relations in the chiral ring. In
particular, the operator $(\tilde Q Q)^2$ is no longer a chiral
primary, and gets a positive anomalous dimension. This is of course
expected since the perturbation by $(\tilde Q Q)^2$ is irrelevant in
the ${\cal N}=3$ theory, as dicussed in the previous subsection.

Consider the example of ${\cal N}=3$ $SU(N)$ theory with 1 adjoint
hypermultiplet $(Q,\tilde Q)$. The superpotential imposes chiral
ring relations of the form \eqn\sutarel{ {\rm Tr}([[\tilde
Q,Q],Q]\cdots)\simeq 0,~~~~~ {\rm Tr}([[\tilde Q,Q],\tilde
Q]\cdots)\simeq 0. } The chiral primaries are given by traces of
strings of $Q$ and $\tilde Q$ symmetrized with respect to
permutation. The trace of a string of $Q$'s and $\tilde Q$'s that is
not symmetrized will in general acquire anomalous dimension, and may
mix with other operators. The spectrum of such operators can be
mapped to that of the Hamiltonian of an $SU(2)_f$ spin chain, as we
will discuss later.

\subsec{${\cal N}=2$ superpotential deformations}

Let us now consider the ${\cal N}=2$ Chern-Simons-matter theory,
with matter fields $\Phi_i$ ($i=1,\cdots,n$) in the representation
$R_i$ of the gauge group $G$. We can deform the theory by a
superpotential \eqn\supage{ W = P(\Phi_1,\cdots,\Phi_n), } $P$ being
a gauge invariant polynomial in the $\Phi_i$'s. We will consider
classically marginal deformations, i.e. $P$ of homogenous degree 4.
We can write \eqn\alapt{ P(\Phi_1,\cdots,\Phi_n) = \sum_{i,j,k,l}
\alpha_{ijkl} (\Phi_i\otimes \Phi_j\otimes \Phi_k\otimes \Phi_l)^G }
where the superscript $G$ means taking the $G$-invariant part of
$R_i\otimes R_j\otimes R_k\otimes R_l$. When there is more than 1
singlet in the tensor product, there are more components of
$\alpha_{ijkl}$, which we suppress here. We would like to know where
the theory flows to under the deformation by \supage\ in the UV.

Firstly, standard non-renormalization arguments forbid other
superpotential terms from being generated. More precisely, if
$\alpha_{ijkl}=0$ (for {\sl all} singlets in $R_i\otimes R_j\otimes
R_k\otimes R_l$) in the classically superpotential, then
$\alpha_{ijkl}$ stays zero in the quantum superpotential. Let us
recall the argument. The theory has a classical $U(1)$ R-symmetry,
which we denote by $U(1)'$, under which $\Phi_i$ has charge $\half$.
We must distinguish this from what we call the quantum $U(1)_R$
symmetry, which is part of the superconformal symmetry of the IR
theory, and the corresponding $U(1)_R$ charges of fields can be
renormalized, as we have seen in the previous section. $U(1)'$ is
nevertheless a good global symmetry of the quantum theory, and
guarantees by holomorphy that the superpotential is quartic in the
$\Phi_i$'s. To proceed, we shall promote $\alpha_{ijkl}$ to a
neutral chiral superfield $Z_{ijkl}$. We have a $U(1)^n$ symmetry,
under which \eqn\astuone{ \Phi_i\to
e^{i\alpha_i}\Phi_i,~~~~~Z_{ijkl}\to
e^{-i(\alpha_i+\alpha_j+\alpha_k+\alpha_l)}Z_{ijkl}. } This
guarantees that the superpotential must be a sum of terms of the
form $Z_{ijkl}(\Phi_i\otimes\Phi_j\otimes\Phi_k\otimes\Phi_l)^G$.
However, when there are more than 1 singlets in the tensor product,
our argument does not forbid other singlets to be generated, even if
they are zero in the classical superpotential.

In supersymmetric Yang-Mills theories, it is usually possible to
further argue non-renormalization of each coupling in the
superpotential (modulo global anomalies say in four dimensions), by
promoting the gauge coupling to a chiral superfield. This is not
possible for the Chern-Simons level $k$, since it would otherwise
violate (small) gauge invariance. So nothing protects the
$\alpha_{ijkl}$'s from $1/k$ corrections, and in fact, they do
receive such corrections.

Suppose all $\alpha_{ijkl}$'s are small. Then the RG flow is
dictated by the anomalous dimension of ${\cal
O}_{ijkl}=(\Phi_i\otimes\Phi_j\otimes\Phi_k\otimes\Phi_l)^G$. This
operator is a chiral primary in the ${\cal N}=2$ CS-matter theory
(with no superpotential), and hence its dimension is determined in
terms of its $U(1)_R$ charge,
$\Delta_{ijkl}=q_i^R+q_j^R+q_k^R+q_l^R$. The quantum corrected
$U(1)_R$ charge $q_i^R$ of $\Phi_i$ appears to be always less than
${1\over 2}$. When $\bigoplus R_i$ is the sum of a (reducible)
representation and its conjugate, the theory can be deformed into
the ${\cal N}=3$ theory and $q_i^R<\half$ follows easily as in the
previous subsections. We computed the two-loop correction to $q^R_i$
explicitly in the theory with $M$ adjoint matter fields in the
planar limit in appendix D.1, and indeed find that the correction is
negative. We expect $q_i^R<\half$ to hold in general.
Consequently, all the quartic
superpotential terms are relevant perturbations, and will take the
${\cal N}=2$ CS-matter theory away from $\alpha_{ijkl}=0$.

On the other hand, if one or several $\alpha_{ijkl}$ are much bigger
than $1/k$, at least when $k\gg 1$ these $\alpha_{ijkl}$'s dominate
and the theory is again approximated by the Wess-Zumino model, which
has positive beta function and hence these $\alpha_{ijkl}$'s
decrease in the IR. In conclusion, $\alpha_{ijkl}$'s are bounded as
the theory flows to the IR.

Therefore, the ${\cal N}=2$ theory deformed by the superpotential
\alapt\ must flow to some other nontrivial IR fixed point at finite
$\alpha_{ijkl}$, which will not be an ${\cal N}=3$ theory in
general. It is easy to write down the two-loop RG equations for the
couplings, \eqn\twologen{ \mu {d\alpha_{ijkl}\over d\mu} =
(q^R_i+q^R_j+q^R_k + q^R_l-2) \alpha_{ijkl}+ {1\over 16\pi^2}({B_i}^r\alpha_{rjkl}+{B_j}^r\alpha_{irkl}+{B_k}^r\alpha_{ijrl}+{B_l}^r\alpha_{ijkr})+{\cal
O}(\alpha^5) } where $B_i^j$ is due to the two loop wave function
renormalization in the Wess-Zumino model with superpotential \alapt\ (i.e. $k\to\infty$ limit),
which is proportional to $\alpha$ and $\overline{\alpha}$.

Specializing to $U(N)$ gauge theory with $M$ adjoint matter fields
$\Phi_i$, and the superpotential deformation \eqn\nust{
W=\sum_{ijkl}\alpha_{ijkl} {\rm Tr}(\Phi_i\Phi_j\Phi_k\Phi_l). } For
simplicity we will work in the planar limit, but the discussion
generalizes easily to the non-planar case as well. We have
${B_i}^j=\half N^2 \alpha_{iklm}\overline{\alpha}^{jklm}$. By $U(M)$
flavor symmetry of the ${\cal N}=2$ theory, we can diagonalize the
Hermitian matrix $B_i^j$ to ${B_i}^j=\half{N^2}c_i\delta_i^j$, with
$c_i\delta_i^j= \alpha_{iklm}\overline{\alpha}^{jklm}$. All the
matter fields $\Phi_i$ have the same renormalized R-charge $q^R$ in
the undeformed theory, and the two-loop RG equation has fixed points
at (assuming $\alpha_{ijkl}\not=0$ for some $j,k,l$)
\eqn\aatwossf{{1\over2}-q^R={N^2\over 32\pi^2} c_i. } Or
equivalently, \eqn\soltaw{ \alpha_{iklm}\overline{\alpha}^{jklm} =
{(\half-q^R)}{32\pi^2\over N^2}\delta_i^j. } Note that the quantum correction to the
R-charge is of order $\lambda^2=(N/k)^2$, and hence the fixed point values of
$\alpha_{ijkl}$ are of order $1/k$. When $M$ is even, this
includes the ${\cal N}=3$ theory with $M/2$ adjoint matters. But
there is clearly more, in fact, a continuous family ${\cal M}$ of
fixed points. The question is whether they survive when higher loop
corrections are included.

While we have not calculated the four-loop beta functions for
$\alpha_{ijkl}$, it appears that there are contributions of the form
${N^4\over k^2}\alpha_{ijmn}\alpha_{klpq}\overline{\alpha}^{mnpq}$,
from supergraphs such as the one in Fig 1. We do not see any
reason why such contributions would cancel. These corrections will
generate nontrivial RG flow along ${\cal M}$.

\vskip 0.5cm \centerline{\vbox{\centerline{
\hbox{\vbox{\offinterlineskip \halign{&#&\strut\hskip0.2cm \hfill
#\hfill\hskip0.2cm\cr \epsfysize=1.1in \epsfbox{loop4.1}  \cr }}}}
{{\bf Figure 1:} a four-loop contribution to the beta function of
$\alpha_{ijkl}$. The dotted lines are ${\cal N}=2$ Chern-Simons
gauge propagators. The solid lines are matter superfield
propagators. }}} \vskip 0.5cm

In general, since the two-loop fixed point values of $\alpha$'s are
of order $1/k$, higher loop corrections will be suppressed by $N/k$.
So at least for weak coupling $N/k\ll 1$, the full RG flow can be
approximated by a flow on the manifold ${\cal M}$. Note that ${\cal
M}$ is compact and smooth, and is freely acted by the $U(M)$ flavor
symmetry. The RG flow can then be represented by the flow according
to a vector field on \eqn\mla{{\cal M}_0={\cal M}/U(M)\simeq
V//U(M),} where $V$ is the parameter space of $\alpha_{ijkl}$'s. For
example, when $M$ is even, the ${\cal N}=3$ theory with $M/2$
adjoints corresponds to one point in ${\cal M}_0$. In general, the
number of critical points (if discrete) of the vector field that
dictates the RG flow, counted with sign, is given by the Euler
characteristic $\chi({\cal M}_0)$. So we learn that if only discrete
RG fixed points (up to the $U(M)$ action) survive when higher loop
corrections are included, their number is at least $\chi({\cal
M}_0)$. In fact, the cohomology of ${\cal M}_0$ is generated by the
Chern classes of the rank-$M$ tautological bundle on ${\cal M}_0$,
and hence $\chi({\cal M}_0)$ is positive and grows with $M$.

The two-loop fixed point locus ${\cal M}$ has codimension $M^2$ in
$V$. The superpotential $W$ imposes chiral ring relations $\partial
W/\partial\Phi_i\sim 0$, and in particular ${\rm Tr}(\Phi^i
{\partial W\over \partial\Phi_j})$ are $M^2$ descendants, while the
remaining quartic chiral operators are in one-to-one correspondence
with chiral primaries. It follows that on ${\cal M}$, the two-loop
renormalization of the $U(1)_R$ charge vanishes. It would be
interesting to know whether/how the $U(1)_R$ charge is renormalized
at the exact conformal fixed points near ${\cal M}$, when higher
loop effects are included.


\newsec{Large $N$ limit}

\subsec{The abelian theory at large $N_f$}

As a starter, we will consider the $U(1)$ Chern-Simons-matter
theories in the large $N_f$ limit, i.e. $k, N_f\to \infty$, with
$\lambda =4\pi N_f/k=q^2N_f$ kept finite (the 't Hooft coupling
$\lambda$ for the nonabelian theory will be defined differently).
This limit of the $U(1)$ theory is rather trivial. For example,
finite dimensional operators only receives anomalous dimension to
subleading order in $1/N_f$. Nevertheless, the theory can have
nontrivial thermodynamics. It is a standard exercise to compute the
free energy of the theory in the infinite $N_f$ limit, by a saddle
point approximation of the path integral. The details of the
calculation can be found in appendix B. We shall discuss the main
results here.

The $U(1)$ ${\cal N}=3$ theory with $N_f$ charged hypermultiplets,
as well as say $U(1)$ ${\cal N}=2$ CS-matter theory with $N_f$ pairs
of oppositely charged matter fields $(Q^i, \tilde Q^i)$, have rather
trivial thermodyanmics at large $N_f$: their free energies at finite
temperature are the same as that of the free field theory, at order
${\cal O}(N_f)$. One can easily convince himself/herself by
examining the cancelation among the Feynman diagrams. It can also be
confirmed through a saddle point analysis, as shown in appendix B.3.
We argued earlier that the ${\cal N}=2$ theory with $(Q,\tilde Q)$
flows to the ${\cal N}=3$ under a superpotential perturbation, and
hence we expect that when $1/N_f$ corrections are included, the free
energy of the former to be greater than the latter, as the number of
degrees of freedom decreases under RG flow, although we have not
checked this explicitly.

The ${\cal N}=2$ $U(1)$ theory with equally charged matter fields,
on the other hand, has nontrivial free energy in the infinite $N_f$
limit. The free energy of the theory in flat space, $F(T) = {1\over
A}\ln Z(T)$ ($A$ being the spatial area), takes the form
\eqn\ccfree{ F(T) = N_f T^2 (c_0(\lambda)+{\cal O}({1/ N_f})) } The
function $c_0(\lambda)$ can be computed from the saddle point
approximation. It is an analytic function in $\lambda$ except at
$\lambda=0$ due to an infrared divergence. This divergence is absent
in the free energy of the theory on a sphere. $c_0(\lambda)$
decreases monotonously as $\lambda$ increases, and asymptotes to a
nonzero value in the $\lambda\to\infty$ limit. At weak 't Hooft
coupling, it is given by \eqn\smallcofr{ c_0(\lambda) =
{7\zeta(3)\over 4\pi} + {\lambda^2 (\ln\lambda)^3\over 32\pi^2}
+\cdots, ~~~~~~~~ \lambda\ll 1. } In the strong coupling limit, we
find $c_0(\infty) \simeq 0.593071$. The function $c_0(\lambda)$ is
plotted (at large coupling) in Figure 2.
 \vskip 0.3cm
\centerline{\vbox{\centerline{ \hbox{\vbox{\offinterlineskip
\halign{&#&\strut\hskip0.2cm \hfill #\hfill\hskip0.2cm\cr
\epsfysize=1.8in \epsfbox{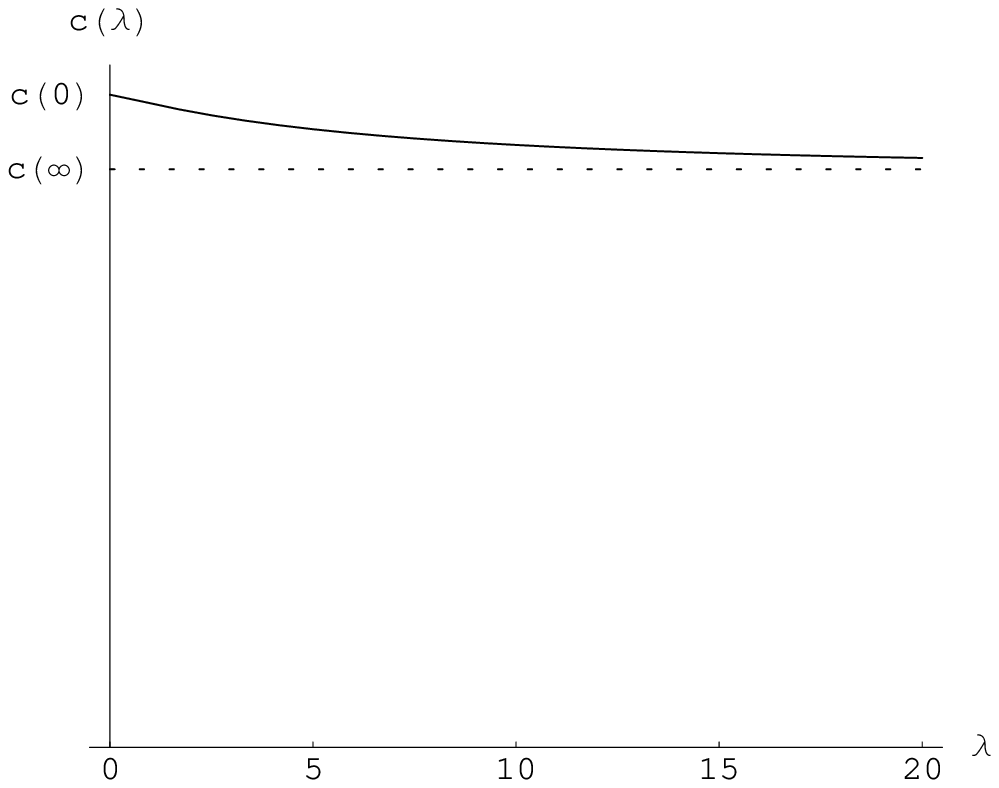}  \cr }}}} \centerline{{\bf
Figure 2:} The free energy as a function of the 't Hooft coupling
$\lambda$.}}} \vskip 0.3cm

More generally, one can consider ${\cal N}=2$ QED with $N_f$ charged
matter fields, with ${\cal N}=2$ Chern-Simons term at level $k$. In
the IR the Yang-Mills coupling is irrelevant, and the theory flows
to the Chern-Simons-matter theory. On the other hand, we can take
the Chern-Simons coupling to infinity ($k/N_f\to 0$), and consider
the strongly coupled IR fixed point of ${\cal N}=2$ QED. The
question is whether the $k/N_f\to 0$ limit and the IR limit are
interchangeable. The answer appears to be yes, as far as the free
energy is concerned. One can compute the low temperature free energy
of ${\cal N}=2$ QED in the large $N_f$ limit by the saddle point
approximation, and arrive at the same result as that of the ${\cal
N}=2$ abelian CS-matter theory at infinite $\lambda$.


It is a much more subtle (and important) question whether the
infinite 't Hooft coupling limit of the $U(N)$ ${\cal N}=2$ or
${\cal N}=3$ CS-matter theory (i.e. $k/N\to 0$) captures the IR
fixed point of corresponding Yang-Mills theory with matter fields.
The answer is not clear to us.

\subsec{The $U(N)$ theory: general remarks on the operator spectrum}

Now let us turn to the $U(N)$ Chern-Simons theory coupled to $N_f$
fundamental matter fields. We will focus on the large $N$ limit,
with $\lambda=N/k$ and $c=N_f/N$ finite. The 't Hooft limit of this
theory is much less trivial than the abelian theory.
We will see that this theory has a number of features suggesting
that it should be dual to a (non-critical) string theory in $AdS_4$.
The radius of the $AdS_4$ scales like $R\sim \sqrt{N} l_{pl}$, and
the string scale is finite compared to $1/R$ when the 't Hooft
coupling is finite.

Let us examine the spectrum of gauge invariant operators, starting
with the {\sl free} theory. In the free theory there are infinitely
many conserved currents of the form \eqn\assphi{ J_{\mu_1\mu_2\cdots
\mu_s}^{\bar i j}=\bar\phi^i D_{(\mu_1}D_{\mu_2}\cdots D_{\mu_s)}
\phi^j + \cdots } where the scalar $\phi$ is in the fundamental
representation of $U(N)$. If there is an $AdS_4$ dual, these
currents should be dual to spin $s$ massless gauge fields in the
bulk in the $\lambda\to 0$ limit \KlebanovJA. The spectrum of gauge
invariant operators in the free theory is analyzed in appendix C by
studying the thermodynamics of the theory on a sphere. An expected
transition from $N_f^2$ degrees of freedom at low temperature to
$NN_f$ degrees of freedom at high temperature is demonstrated
explicitly.

 In the interacting
theory most of \assphi\ will acquire anomalous dimensions. There
will still be conserved currents, the stress-energy tensor
$T_{\mu\nu}$, the $U(1)_R$ current $J^R_\mu$, and the flavor current
$J^{U(N_f)}_{\mu\nu}$, as well as fermionic currents related by
supersymmetry. These are dual to the graviton, graviphoton, and
$U(N_f)$ gauge fields in the bulk.

In the ${\cal N}=2$ theory with fundamental matter, there are no
chiral primaries of small dimensions, since gauge invariant
operators of dimension less than $N$ must involve matter fields in
both fundamental and anti-fundamental representations, and hence
must depend on both $\phi$ and $\bar\phi$. This suggests the absence
of Kluza-Klein modes in the bulk. When $N_f\geq N$, there are
baryonic chiral primary operators \bartoo.\foot{In the case of
$U(N)$ gauge group with $N_f$ fundamental matter, the $U(1)$ part
only contribute to the anomalous dimensions to subleading order in
$1/N_f$. } The bulk theory has 8 supersymmetries, and in particular
the $U(N_f)$ gauge fields should lie in four dimensional ${\cal
N}=2$ vector multiplets. The latter also contains a complex scalar,
which is dual to the operators \eqn\scantwo{ \bar\phi^i
\phi_j,~~~~~\bar\psi^i \psi_j + {8\pi\over k}(\bar\phi^l\phi_l)
(\bar\phi^i\phi_j). } By examining their supersymmetry variations,
one can see that they lie in the same supermultiplet as the $U(N_f)$
flavor current. A surprising consequence is that the dimension of
$\bar\phi^a \phi_b$ is protected to be 1, despite that it is not a
chiral primary operator. This is checked explicitly at two-loop in
appendix D.1. Recall that in the ${\cal N}=2$ theory, the chiral
primaries in fact have negative anomalous dimensions due to the
renormalization of $U(1)_R$ charge.

In the ${\cal N}=3$ theory there are in addition mesons \KapustinHA,
${M^i}_j=\tilde Q^i_a Q_j^a$, which are dual to a complex scalar in
the ${\cal N}=4$ $U(N_f)$ gauge multiplet in $AdS_4$. There are two
more $U(N_f)$ adjoint complex scalars in the four dimensional ${\cal
N}=4$ gauge multiplet. One of them is $\tilde \psi^i
\psi_j+{8\pi\over k}(\tilde Q^l Q_l)\tilde Q^i Q_j$, the other one
consists of $\overline Q^i Q_j-\tilde Q^i \overline{\tilde Q}_j$ and
$\overline\psi^i \psi_j-\tilde \psi^i \overline{\tilde\psi}_j +
\cdots$.

In ${\cal N}=2$ or ${\cal N}=3$ theories with several {\sl adjoint}
matter fields, there are a lot more chiral primaries given by
unsymmetrized traces of chiral fields, as described before. Their
number grows exponentially in dimension. It would be interesting to
understand its meaning in the holographic dual (if exists). However,
this behavior is likely to be ``non-generic", i.e the ${\cal N}=2$
theory deformed by a generic superpotential will flow to an IR fixed
point, where the number of chiral primaries may not have the
exponential growth.


Since the Chern-Simons gauge field does not carry propagating
degrees of freedom, the central charge of the theory with $N_f$ fundamental flavors is
expected to be of the form \eqn\centaa{ c(\lambda) =NN_f f(\lambda),
} where $f(\lambda)$ is some nonzero function, at least for sufficiently weak (but finite) 't Hooft
coupling $\lambda$. Comparing this to the central charge expected
for a theory of gravity in $AdS_4$, of radius $R\gg l_{pl}$, $c\sim
(R/l_{pl})^2$, we expect \eqn\plna{ R\sim (NN_f)^\half
f(\lambda)^\half l_{pl} } When $N,N_f$ are large, we expect the
$AdS_4$ dual to be a theory of gravity coupled to $U(N_f)$ gauge
fields as well as massive fields, with large $AdS_4$ radius
in Planck units. In the next section we will present evidence for
the existence of semi-classical strings in this theory.

\subsec{Twist-1 operators}

Let us now consider the twist-1 operators of the form \eqn\tistw{
{({\cal O}_{n,\Delta})_i}^{j} = \bar\phi_i D_{(\mu_1}\cdots
D_{\mu_n)|} \phi^j\Delta^{\mu_1}\cdots\Delta^{\mu_n} } for large
spin $n$, where $\Delta$ is an arbitrary null vector. The subscript
$|$ indicates the subtraction of traces in Lorentz indices. The
analog of such operators in four dimensional ${\cal N}=4$
super-Yang-Mills theory in the large spin limit is dual to a folded
semi-classical spinning string in $AdS_5$. We anticipate a similar
interpretation in CS-matter theory: \tistw\ should be dual to a long
spinning open string in $AdS_4$ in the limit of large $n$. As noted
in \GubserTV, the operators \tistw\ are closely related to
light-like open Wilson lines, \eqn\openwil{ \bar\phi_i(x) {\rm
P}e^{i\int_y^x A}\phi^j(y) = \sum_{n=0}^\infty {1\over n!}{({\cal
O}_{\mu_1\cdots\mu_n})_i}^j (x-y)^{\mu_1}\cdots (x-y)^{\mu_n}. } The
one-loop anomalous dimension of \tistw\ vanishes trivially due to
kinematics. There is a (regularization dependent) 1-loop shift of
the Chern-Simons level $k$, and hence the 't Hooft coupling
$\lambda=N/k$. Since there is no logarithmic divergence at 1-loop,
this shift does not affect the anomalous dimension at 2-loop. One
may need a more careful treatment of the regularization of the
Chern-Simons propagator at higher than two loops.

There is another set of twist-1 operators, of the form $\bar\psi
\gamma_{(\mu_1}D_{\mu_2}\cdots D_{\mu_n)|}\psi$, which may mix with
\tistw, through diagrams such as the one in Figure 3.

\vskip 0.5cm \centerline{\vbox{\centerline{
\hbox{\vbox{\offinterlineskip \halign{&#&\strut\hskip0.2cm \hfill
#\hfill\hskip0.2cm\cr   \epsfysize=.6in \epsfbox{mixing.1}  \cr }}}}
{{\bf Figure 3:} mixing of ${\cal O}_{n,\Delta}$ with $\bar\psi
\gamma_{(\mu_1} D_{\mu_2}\cdots D_{\mu_n)}\psi$. The solid lines are
$\phi$ propagators and the double lines are $\psi$ propagators.}}}
\vskip 0.5cm

Such
contribution will however be suppressed at large spin $n$. For
example, the loop integral involved in figure 3 takes the form
\eqn\mixsupr{ \int d^3k d^3l {(k\cdot\Delta)^n {\slash\!\! l}\over
(k^2)^2 l^2 (k+l-p)^2} \to \int d^3k {(k\cdot\Delta)^n
({\slash\!\!\! k}-{\slash\!\!\! p})\over (k^2)^2 |k-p|} }
After introducing a Feynman parameter and performing the integral over
$k$, one ends up with \eqn\asinggg{ \sim(p\cdot
\Delta)^{n-1}{\slash\!\!\!\!\Delta}\ln\Lambda\int_0^1 dx \,n(1-x)
x^{n-{3\over 2}}, } whose contribution to the operator mixing goes
like $1/n$ at large $n$. In conclusion, we can ignore operator
mixing in our computation.

We will now describe the computation of the leading two-loop anomalous
dimension of \tistw. The leading contribution to the anomalous
dimension comes from 2-loop diagrams as in figure 4, $(a)-(d)$.

\vskip 0.5cm
\centerline{\vbox{
\hbox{\vbox{\offinterlineskip
\halign{&#&\strut\hskip0.2cm
\hfill
#\hfill\hskip0.2cm\cr
 ~~~~~~~~~~~&  \epsfysize=0.6in \epsfbox{longops.1} &~~~~~~~& \epsfysize=0.6in \epsfbox{longops.2} &~~~~~~~& \epsfysize=0.6in \epsfbox{longops.3}
 &~~~~~~~&
  \epsfysize=0.6in \epsfbox{longops.4} & \cr &
  $(a)$ && $(b)$ && $(c)$ && $(d)$ & \cr
}}} \centerline{{\bf Figure 4:} two-loop contributions to the anomalous dimension of $J_{\mu_1\cdots\mu_n}$.}}}
\vskip 0.5cm

The diagram $(d)$ includes the 1-loop corrections to the gauge field
propagator from a loop of the gauge field, ghost, and the matter
fields. The gauge and ghost bubbles cancel, and only the matter
field bubble contributes. These diagrams are computed in appendix
D.2. We find that $(b)$ vanishes,\foot{More generally, whenever two
adjacent gauge field propagators coming out of ${\cal O}$ are joined
by a cubic Chern-Simons vertex, the diagram is zero.} $(a)$ and
$(c)$ give contributions of order $\lambda^2$ that is finite in the
$n\to\infty$ limit, whereas $(d)$ gives a contribution of order
$\lambda^2 N_f/N$ that grows like $\ln(n)$ in the large $n$ limit.
Diagrams that involve exchange of gauge fields or matter fields
between the two external scalar lines, such as the ones in figure 5,
are suppressed in the large $n$ limit.

\vskip 0.5cm
\centerline{\vbox{
\hbox{\vbox{\offinterlineskip
\halign{&#&\strut\hskip0.2cm
\hfill
#\hfill\hskip0.2cm\cr
 ~~~~~~~~~~~~~~~~~~~~~&  \epsfysize=0.6in \epsfbox{longops.5} &~~~~~~~& \epsfysize=0.6in \epsfbox{longops.6} &~~~~~~~& \epsfysize=0.6in \epsfbox{longops.7}
 & & \cr &
  $(e)$ && $(f)$ && $(g)$ & \cr
}}} \centerline{{\bf Figure 5:} some diagrams that are suppressed in the large spin limit. }}}
\vskip 0.5cm

We find that the anomalous dimension of the operator
$J_{\mu_1\cdots\mu_n}$ is given to two-loop order by \eqn\amon{
\Delta-n-1 = const\cdot{\lambda^2}+ {N_f\over N}{\lambda^2\over
8}\ln(n)+ {\rm higher~order~corrections}, ~~~~~~n\gg 1. } Note that
the above two-loop computation is essentially identical for ${\cal
N}=3$ theories, as well as theories with adjoint matter fields (for
which the coefficient of $\ln(n)$ does not depend on $N$), since the
vertices involving only matter fields do not contribute at this
order.

Naively one might have expected higher loop corrections to give
contributions to the anomalous dimension that grow like powers of
$\ln(n)$. On the other hand, we have already seen nontrivial
cancelations: one might have expected diagram $(a)$ in Figure 4 to
grow with $n$, but in fact it doesn't. In four dimensional gauge
theories, the $\ln(n)$ growth of the anomalous dimension of twist-2
operators at large spin $n$ is well known, and is related to the
cusp anomalous dimension of Wilson lines
\refs{\KorchemskyWG,\KorchemskyXV,\KruczenskiFB}. Namely, the Wilson
line consisting of two straight pieces with a turn of (Lorentzian)
angle $\gamma$ acquires an anomalous dimension of the form
$\Gamma_{\rm cusp}(\gamma,\lambda)=\gamma \Gamma_{\rm
cusp}(\lambda)+{\cal O}(\gamma^0)$ in the large $\gamma$ limit,
where $\Gamma_{\rm cusp}(\lambda)$, the cusp anomalous dimension, is
a function of the coupling only. The coefficient of $\ln(n)$ in the
anomalous dimension of the twist-2 operator is in fact equal to
$-2\Gamma_{cusp}(\lambda)$. The linearity of $\Gamma_{\rm
cusp}(\gamma,\lambda)$ in $\gamma$ at large angles is argued in
\KorchemskyWG\ by examining in a physical gauge the logarithmic
divergences from the integration over momenta collinear to the
(almost) light-like direction \EllisTY\ along an edge of the Wilson
line.

In Chern-Simons-matter theory in three dimensions, the
matter-corrected gauge propagator is the sum of a piece that takes
the same form as the Yang-Mills propagator in ${\sl four}$
dimensions in ${\sl position}$ space, i.e. proportional to
$\eta_{\mu\nu}(x-y)^{-2}$ plus gauge dependent terms, and the pure
Chern-Simons propagator, which in a physical gauge is a contact term
in position space.\foot{In the axial gauge, for instance, the pure
CS propagator is given by $\epsilon_{ij}{\rm
sgn}(x_3-y_3)\delta^2(\vec{x}-\vec{y})$.} In fact, the YM-like piece
of the matter-corrected gauge propagator in diagram $4(d)$ is
responsible for the $\ln(n)$ growth of the two-loop anomalous
dimension computed above. It is therefore plausible that a scaling
argument similar to that of \KorchemskyWG\ should go through for
Chern-Simons-matter theory as well. Hence we anticipate the $\ln(n)$
growth of the anomalous dimension of twist-1 operators to hold to
all orders in perturbation theory.\foot{Although, we do not
understand the possible subtleties in the scaling argument involving
the singular looking CS propagator in a physical gauge, which
clearly deserves a more careful analysis.}

When $N_f/N\ll 1$, we expect the large $n$ growth of the anomalous
dimension to be of the form ${N_f\over N}f(\lambda) \ln(n)$, for
some function $f$. In \GubserTV\ a classical folded closed string
spinning in AdS space with energy $E$ and spin $J$ was considered,
and it was found that in the large spin limit, \eqn\srrs{ E - J =
{R^2\over \pi\alpha'} \ln ({\alpha'J\over R^2})+\cdots } The result
for a spinning open string is similar, with $\alpha'$ replaced by
$2\alpha'$. If the ${\cal N}=2$ $U(N)$ Chern-Simons-matter theory is
dual to a string theory in $AdS_4$, we expect (for $N_f/N\ll 1$)
\eqn\exptstr{ {R^2\over\alpha'} \sim f(\lambda){N_f\over N} } On the
other hand, by comparing the central charge in gravity with that of
the gauge theory, we expect \eqn\srrerr{ {R} \sim (NN_f)^\half
\tilde f(\lambda) l_{pl} } for some other function $\tilde
f(\lambda)$. Therefore, \eqn\tfplk{ {\alpha'\over l_{pl}^2} \sim
N^2{\tilde f(\lambda)^2\over f(\lambda)} } This is consistent with
the expectation (say by examining three-point functions) that in the
large $N$ limit, the string coupling is independent of $N_f$.

Note that the $U(N_f)$ gauge fields in the bulk also has coupling of
order $1/N$ (times some function of $\lambda$), and hence
(four-dimensional) 't Hooft coupling of order $N_f/N$. In the large
$N$ limit, with $N_f/N=c\ll 1$, the bulk theory is effectively
weakly coupled, which justifies the above discussion. A naive
attempt to get large $AdS_4$ radius in string units is to take
$N_f/N$ large. In this case, however, the bulk gauge theory becomes
strongly coupled. This is particularly intriguing in the case of
${\cal N}=3$ CS-matter theory, where the bulk theory contains
four-dimensional $U(N_f)$ ${\cal N}=4$ SYM coupled to ${\cal N}=3$
supergravity, as well as infinitely many massive fields. As $N_f/N$
increases, we expect the radius $R$ to increase as well. In the
limit of large $N_f/N$, the suitable description of the bulk theory
may involve a further holographic dual to a five dimensional theory
of gravity. This suggests that flavor singlets in the
Chern-Simons-matter theory of the form \eqn\sinfl{ (\tilde Q^{i_1}
Q_{i_2}) (\tilde Q^{i_2} Q_{i_3})\cdots (\tilde Q^{i_n} Q_{i_1}) }
could be dual to closed strings in the five dimensional theory.

\subsec{Supersymmetric Wilson loops}

In this section we will describe supersymmetric Wilson loops in the
theory. First consider the ${\cal N}=2$ Chern-Simons theory. The
supersymmetry transformations of the gauge field $A_\mu$ and
auxiliary field $\sigma$ are of the form \eqn\susyntwo{
\delta_\epsilon A_\mu = {i\over 2}(\bar\varepsilon \gamma_\mu \chi -
\bar\chi \gamma_\mu\varepsilon), ~~~~~\delta_\epsilon \sigma =
{i\over 2} (\bar\varepsilon \chi - \bar\chi \varepsilon). } The
Wilson line \eqn\wila{ P \exp\left[\int d\tau\left( A_\mu \dot x^\mu
+ \sigma |\dot {\vec x}| \right) \right] } locally preserves $1/2$
supersymmetries, whose supersymmetry parameters are solutions of
$\gamma_\mu {\dot x^\mu\over |\dot{\vec
x}|}\varepsilon=-\varepsilon$. Globally, only the straight Wilson
line can preserve half of the ${\cal N}=2$ supersymmetries.
Similarly, the straight Wilson line also preserves one half of the
special supercharges of the superconformal algebra $OSp(2|4)$. More
generally, any conformal transformation of the straight Wilson line
\wila\ will preserve one half of the supersymmetries as well. As
pointed out in \DrukkerRR, large conformal transformations in ${\bf
R}^3$ that take a point on the Wilson line at finite distance to
infinity will not preserve the expectation value of the BPS Wilson
line.

In the pure Chern-Simons theory, it is well known that the Wilson
loops are topological \WittenHF. This is no longer the case when the
Chern-Simons theory is coupled to matter fields. Note that the gauge
field propagator (in Feynman gauge) in the 1PI effective action
takes the form \eqn\aassprop{ {\delta^{ab}\over
1-\Pi(k)^2}\left[{\epsilon_{\mu\nu\rho}p^\rho\over
2p^2}+\Pi(k){\delta_{\mu\nu}-p_\mu p_\nu/p^2\over p }\right] } where
$\Pi(k)$ is due to loops involving the matter fields, $a, b$ are
gauge indices. $k$ here is the Chern-Simons level, not to be
confused with momentum. The structure of \aassprop\ is constrained
by conformal invariance.
 From the one-loop matter bubble correction we have
$\Pi(k)= -{\pi N_f\over 4k}+{\cal O}({1\over k^2})$. The Fourier
transform of \aassprop\ to position space is \eqn\postiapo{ \langle
A_\mu^a(x) A_\nu^b(y)\rangle \sim {{\delta_a}^b\over
1-\Pi(k)^2}\left[{\epsilon_{\mu\nu\rho}(x-y)^\rho\over
2|x-y|^3}+\Pi(k) \left({\delta_{\mu\nu}\over |x-y|^2 }-\partial_\mu
\partial_\nu \ln|x-y|\right)\right] } We see that the second term on
the RHS of \postiapo\ takes the same form as the Yang-Mills
propagator in {\sl four} dimensions. Using the method of \DrukkerRR,
it may be possible to compute say the exact expectation value of
circular Wilson loops by localizing the contributions to a point.
(See \vandeWeteringXP\ for a perturbative approach to Wilson loops
in pure CS theory.) We leave this problem to future investigation.

There are also supersymmetric Wilson lines in the ${\cal N}=3$ theory, of the form
\eqn\nthreewil{ P\exp\left[ \int d\tau \left( A_\mu \dot x^\mu + s_i \dot y^i \right) \right] }
where $s_i$'s are the $SU(2)_R$ multiplet of auxiliary fields containing $\sigma$ and $\Phi$. $(x^\mu(\tau), y^i(\tau))$
parameterizes a path in ${\bf R}^3\times {\bf R}^3$, where the second ${\bf R}^3$ is an internal space
acted by the $SU(2)_R$. The condition for \nthreewil\ to locally preserve supersymmetries is
$|\dot {\vec x}|=|\dot {\vec y}|$. The preserved supersymmetry parameter $\varepsilon_i$ are solutions of
\eqn\eloc{ \gamma_\mu \dot x^\mu \varepsilon_i+i\epsilon_{ijk}\dot y^j \varepsilon_k=0. }
A straight Wilson line of the form \nthreewil\ with constant $\dot y^i$
 preserves $1/3$ of the ${\cal N}=3$ supercharges.

\subsec{The anomalous dimension of a Wilson line with an angle}

As recalled earlier, the $\ln(n)$ growth of the anomalous dimension
of the twist-1 operator is related to the anomalous dimension of a
Wilson line with a turn of angle $\gamma$, in the large $\gamma$
limit. In this subsection we describe an intuitive physical picture
of the latter, following \BelitskyYS. Let us first consider the
Euclidean version. Take a Wilson line consisting of two straight
pieces, labelled by vectors $u,v$, joined at the origin. The turning
angle $\theta$ is given by $\cos\theta={u\cdot v\over |u||v|}$.
Since the Chern-Simons-matter theory is conformal, we can map the
configuration conformally to one on $S^2\times {\bf R}$, with the
two straight pieces of the Wilson line at two points separated by
the angle $\pi-\theta$ on the $S^2$, extending in ${\bf R}$. The
anomalous dimension of the Wilson line is equivalent to the
potential energy between a pair of quark and anti-quark on the
$S^2$.

To leading order the potential energy is determined by the two point
function of the gauge fields \postiapo. The part that contributes is
similar to the four-dimensional Yang-Mills propagator in position
space, resulting in a potential energy \eqn\apotew{ V(\pi-\theta)
\propto \theta \cot\theta + const } Note that this should be
interpreted as the potential due to a quark at $\theta=\pi$ and a
uniform background charge on the $S^2$ that cancels the charge of
the quark, due to the Gauss law constraint.

In Lorentzian signature, the angle is $\gamma=i\theta$, with
$\cosh\gamma={u\cdot v\over |u| |v|}$. The analytic continuation of
the above conformal transformation brings ${\bf R}^{2,1}$ to
$AdS_2\times {\bf R}$, with the time direction of the $AdS_2$
compactified on a circle. The anomalous dimension of the Wilson
loop, to leading order (one-loop in the matter-corrected gauge
propagator), is given by \eqn\afv{ \Gamma(\gamma)\sim
\gamma\coth\gamma + const } Indeed, at large angle $\gamma$, the
anomalous dimension is linear in $\gamma$. This essentially
reproduces the result of the two-loop calculation of section 3.3.
Less obviously, we expect the linearity at large $\gamma$ to hold
for the analytic continuation of the potential function on the
$S^2$, to all orders in perturbation theory, based on arguments
along the lines of \KorchemskyWG.

\subsec{The ${\cal N}=3$ spin chain}

A powerful technique that can be used to understand the operator
spectrum is to map the dilatation operator to a spin chain
Hamiltonian (see \BeisertRY\ for a review of the subject). In this
subsection we consider the ${\cal N}=3$ theory with one adjoint
hypermultiplet matter $(Q,\tilde Q)$. A basic chiral primary
operator is ${\rm Tr}Q^J$. By acting on it with $SU(2)_R\times
SU(2)_f$ symmetry, we obtain a more general class of protected
operators, \eqn\aamor{ {\rm Tr}\left[(v^A Q_A(u))^J \right], } where
$Q_A(u)$ are defined as in appendix A, $u^a$ and $v^A$ are doublets
of $SU(2)_R$ and $SU(2)_f$ respectively. Among these there are
symmetrized traces of $Q$'s mixed with $\tilde Q$'s, which are
${\cal N}=2$ chiral primaries. A perhaps less obvious class of
protected operators are the symmetrized traces of $Q$'s mixed with
$\overline{\tilde Q}$'s.

The operators obtained by inserting a few $\tilde Q$ ``impurities"
in ${\rm Tr}(QQ\cdots Q)$, without symmetrization inside the trace,
is not a chiral primary and will receive anomalous dimension, due to
the superpotential $W={\pi\over k}{\rm Tr}([Q,\tilde Q]^2)$.
Similarly, if $\overline{\tilde Q}$'s are inserted in ${\rm Tr}Q^J$,
but not symmetrized, the operator is not protected and has anomalous
dimension. We shall compute the two-loop anomalous dimension of such
operators in the planar limit, assuming that the operator is very
long and the impurities are far away from one another. The kind of
diagrams that contribute are shown in Figure 6.

\vskip 0.5cm \centerline{\vbox{ \hbox{
\centerline{\vbox{\offinterlineskip \halign{&#&\strut\hskip0.2cm
\hfill #\hfill\hskip0.2cm\cr & \epsfysize=0.8in
\epsfbox{spinchain.1}
 &~~~~~~~~~~~& \epsfysize=0.8in \epsfbox{spinchain.2}
 & & \cr  &
  $(a)$ && $(b)$ & \cr
}}}} \centerline{ {\bf Figure 6.}}}} \vskip 0.5cm

Let us first consider $\tilde Q$-impurities, corresponding to
excitations of the $SU(2)_f$ spin chain, since $Q$ and $\tilde Q$
are a doublet of $SU(2)_f$. The operators of interest are of the
form
$$
{\rm Tr}(Q\cdots Q\tilde QQ\cdots Q\tilde Q Q\cdots Q)
$$
Diagram $6(a)$ may involve a sextic scalar coupling coming from the
superpotential, of the form \eqn\supte{ {4\pi^2\over k^2} {\rm Tr}
\left( [\tilde Q,[\tilde Q, Q]] \overline{[\tilde Q,[\tilde Q, Q]]}
+ [Q,[\tilde Q, Q]] \overline{[ Q,[\tilde Q, Q]]} \right) } There
are also sextic scalar interactions coming from the coupling to
${\cal N}=2$ Chern-Simons gauge multiplet, ${\rm Tr}([\overline
Q,\sigma][\sigma,Q]+[\overline{\tilde Q},\sigma][\sigma,\tilde Q])$,
but they do not in fact contribute to the anomalous dimension in the
planar limit. Similarly, the diagrams involving fermions (Fig 6
$(b)$) do not contribute in the planar limit either. It is a simple
exercise to derive the two-loop spin chain Hamiltonian from \supte,
and we find \eqn\shchah{ H^{SU(2)_f}_{(2)} = {\lambda^2\over
4}\sum_i (4P_{i,i+1}-P_{i,i+2}-6), } where $P_{i,j}$ is the operator
that interchanges the $i$-th and $j$-th sites. This is the
Hamiltonian of an $su(2)$ XXX spin-$\half$ chain with
next-to-nearest neighbor interactions. Such a spin chain is in fact
{\sl not} integrable. This suggests that the dual string worldsheet
sigma model, if exists, will probably not be integrable either.

Let us now consider $\overline{\tilde Q}$ impurities, corresponding
to excitations of the $SU(2)_R$ spin chain. In this case, diagram
$(a)$ in Fig 6 gives rise to next-to-nearest neighbor interactions,
$\sum P_{i,i+2}$, whereas diagram $(b)$ gives rise to nearest
neighbor interactions, $\sum P_{i,i+1}$. The resulting two-loop spin
chain Hamiltonian $H_{(2)}^{SU(2)_R}$ in fact takes the identical
form as \shchah.

It follows from \shchah\ that the spectrum of anomalous dimensions
for operators in the $SU(2)_f$ or $SU(2)_R$ sector, in the limit of
large $J$ (length) and few impurities, is given by \eqn\anoma{
\Delta-{J\over 2}={\lambda^2\over 4}\sum_m\left[2\sin({\pi l_m\over
J})\right]^4+{\cal O}(\lambda^3),~~~~l_m\in {\bf Z}, } where $l_m$
is the ``momentum" of the $m$-th impurity. This is clearly a very
different dispersion relation from that of string modes in ordinary
pp-wave backgrounds \BerensteinJQ.

In the $SU(2)_f\times SU(2)_R$ invariant notation of appendix A,
$Q=q^{11}$, $\tilde Q=q^{21}$, and $\overline{\tilde Q}=-q^{12}$.
There are two more basic impurities, given by insertions of fermions
$\psi^{11}_\alpha$. As we will partially justify, insertions of
other fields such as $\overline{Q}=q^{22}$ in the infinitely long
spin chain can be thought of as bound states of the four basic
impurities. The superconformal group $OSp(3|4)$ has a subgroup
$SU(1|2)$ that leaves $q^{11}$ invariant. The bosonic part of
$SU(1|2)$, $U(1)\times SU(2)$, is generated by $\Delta-J_R^3$ and
the rotations in ${\bf R}^3$. The fermionic generators are the
chiral supercharges $Q_\alpha$ and $\overline S^\alpha$.

The symmetrized traces of $Q$'s with $\tilde Q$ or $\overline{\tilde
Q}$ insertions, i.e. impurities with zero momenta, are protected
because they sit in reduced representations of $SU(1|2)$. In fact,
at zero momentum, $q^{21}$ is a singlet, whereas
$(q^{12},\psi^{11}_\alpha\sim Q_\alpha q^{12})$ form a fundamental
representation of $SU(1|2)$. A more general representation of
$SU(1|2)$ can be obtained by acting $Q_\alpha$ on a primary of
anomalous dimension and spin $(h,j)$. The resulting representation
consists of $U(1)\times SU(2)$ content \eqn\suotw{ (h,j)\oplus
(h+\half,j+\half)\oplus (h+\half,j-\half)\oplus (h+1,j+1) }

It is convenient to consider impurities in an infinitely long spin
chain of $q^{11}$'s, to allow states with a single impurity of
momentum $p$ (finite traces always have total momentum zero by
cyclicity). When there is nonzero momentum, the basic impurities are
no longer in reduced representations of $SU(1|2)$, and will acquire
anomalous dimensions. Due to the superpotential of the ${\cal N}=3$
theory, we have \eqn\astis{ Q_\alpha \psi_\beta^{11}\sim
\epsilon_{\alpha\beta} {1\over k} [[q^{11},q^{21}],q^{11}] } In an
infinitely long spin chain, the extra factors of $q^{11}$ are
unimportant, but the different terms in the commutators receive
different phase factors $e^{2\pi ipn}$. Therefore, we have
\eqn\centralss{ Q_\alpha |\psi^{11}_\beta(p)\rangle \sim
\epsilon_{\alpha\beta} {1\over k} \left[2\sin({\pi p})\right]^2
|q^{21}(p)\rangle } Consequently, the impurities
$(q^{12},q^{21},\psi^{11}_\alpha)$ with nonzero momentum $p$ sit in
a long multiplet of $SU(1|2)$. A somewhat unexpected consequence is
that the $SU(2)_R$ impurity and the $SU(2)_f$ impurity have the same
anomalous dimension, agreeing with our calculations of the two-loop
spin chain Hamiltonian.

Unlike ${\cal N}=4$ SYM, our basic impurities do not sit in any
shortened representation of the supergroup, hence we cannot deduce
the form of the exact anomalous dimensions to all order based on the
deformed superalgebra for the infinite chain. Also note that the
product of two representations corresponding to the basic impurities
is a sum of irreducible representations (as opposed to one long
representation in the case of ${\cal N}=4$ SYM), the two impurity
scattering S-matrix will be determined by several scattering phases.
A detailed exploration of these topics is beyond the scope of the
current paper.

\subsec{The ${\cal N}=2$ spin chain}

Let us briefly describe the spin chain in the ${\cal N}=2$ theory
with one adjoint matter $\Phi=(\phi,\psi)$. A basic chiral primary
is ${\rm Tr}\phi^J$. We will again consider infinitely long chain,
with $SU(1|2)$ symmetry. The basic impurities are $(\bar\phi,
\psi_\alpha, \bar\psi_\alpha,D_\mu)$, forming a long multiplet of
$SU(1|2)$ (unlike the ${\cal N}=3$ case, where such impurities are
bound states of more elementary impurities). The action of the
chiral supercharge $Q_\alpha$ on the fields are schematically
\eqn\actiaq{ \eqalign{&Q_\alpha \bar\phi \sim \bar\psi_\alpha, \cr
&Q_\alpha \bar\psi_\beta =0, \cr & Q_\alpha \psi_\beta \sim
-i\gamma^\mu D_\mu\phi + {2\pi\over k}[[\phi,\bar\phi],\phi],\cr &
Q_\alpha D_\mu \phi \sim {(\gamma_\mu)_\alpha}^\beta {2\pi\over k}
[\phi,\bar\psi_\beta], } } or in terms of impurities with momentum
$p$, \eqn\asimpu{\eqalign{&Q_\alpha |\bar\phi(p)\rangle \sim
|\bar\psi_\alpha(p)\rangle, \cr &Q_\alpha |\bar\psi_\beta(p)\rangle
=0, \cr & Q_\alpha |\psi_\beta(p)\rangle \sim -i\gamma^\mu
|D_\mu(p)\rangle + {2\pi\over k}\left[2\sin(\pi p)\right]^2
|\bar\phi(p)\rangle, \cr & Q_\alpha |D_\mu(p)\rangle \sim
{(\gamma_\mu)_\alpha}^\beta {2\pi\over k}(e^{2\pi ip}-1)|\psi_\beta
(p)\rangle, }} One should be cautious that $D_\mu$ at zero momentum
always have dimension 1, whereas $\phi$ itself has dimension $q^R$,
renormalized to be less than $1/2$.

The two-loop Hamiltonian of the chain made of $\phi$'s and
$\bar\phi$'s takes the form \eqn\taoaj{ H_{(2)}^{{\cal N}=2} =
{\lambda^2\over 4} \sum_i (3P_{i,i+1}-P_{i,i+2}-4)+({\rm
zero~momentum~contribution}) } where the zero momentum contribution
gives the anomalous of the symmetrized traces. Note that in the
continuum limit, the dispersion relation of the $\bar\phi$ impurity
goes like $\sin^2(\pi p)$, as familiar in ordinary pp-wave limits,
in contrast to the $\sin^4(\pi p)$ behavior we saw for the ${\cal
N}=3$ theory.

As discussed in section 2, one can turn on a superpotential
$W=\alpha {\rm Tr} \Phi^4$, and flow to a different SCFT. This
superpotential will kill all the chiral primaries of the form ${\rm
Tr}\phi^J$ except ${\rm Tr}\phi^2$. Together with its descendants
${\rm Tr}\phi\psi_\alpha$ and ${\rm Tr}\psi_\alpha\psi_\beta
\epsilon^{\alpha\beta}$, they are dual to an ${\cal N}=2$
hypermultiplet in the bulk $AdS_4$.

\newsec{Some preliminary attempts to construct the holographic dual}

There are many ways of engineering three dimensional gauge theories
with Chern-Simons coupling in string theory. We will describe some
examples in this section, in attempt to find gravity duals of the
Chern-Simons-matter system. However we have not been able to find
$AdS_4$ dual in the supergravity regime.

\subsec{${\cal N}=2$ $U(N)$ theory with one adjoint}

One can obtain ${\cal N}=2$ $U(N)$ Chern-Simons theory at level $k$
coupled to an adjoint matter as the low energy limit of the world
volume theory of $N$ M5-branes wrapped on a special Lagrangian lens
space $S^3/{\bf Z}_k$ in a Calabi-Yau 3-fold (say the cotangent
bundle of $S^3/{\bf Z}_k$), at least for large $k$. The M-theory
reduces to type IIA string theory compactified on a five-manifold
involving an $S^2$, with $k$ units of $F^{RR}_{(2)}$ flux on the
$S^2$, and the M5-branes turn into D4-branes wrapped on the $S^2$.
The RR-flux on the $S^2$ induces Chern-Simons coupling in the
D4-brane world volume gauge theory. The gauge theory is three
dimensional ${\cal N}=2$ super-Yang-Mills with Chern-Simons
coupling, coupled to an adjoint matter (corresponding to the 2
transverse coordinates of the D4-brane in ${\bf R}^{1,4}$), with no
superpotential. At low energies the Yang-Mills coupling becomes
irrelevant, and the theory flows to ${\cal N}=2$ Chern-Simons-matter
theory.

The gravity solutions of supersymmetric M5-branes wrapped on special
Lagrangian cycles were solved in \GauntlettNG, with the sLag 3-cycle
being either a quotient of $S^3$ or a quotient $H_3$ of the
hyperbolic 3-space. It turns out that a smooth near horizon $AdS_4$
regime in the supergravity exists only when the M5-branes are
wrapped on $H_3$, and not $S^3$ (nor its quotient). Since $H_3$
cannot be realized as a circle fibration over a Riemann surface, the
corresponding M5-brane configuration is ``intrinsically
M-theoretic". In conclusion we do not find an $AdS_4$ dual to
Chern-Simons theory with one adjoint matter in the supergravity
regime. It is plausible, however, that when $\alpha'$-corrections
are included, there could be a smooth near-horizon limit of the
$S^2$-wrapped D4-brane, involving an $AdS_4$ whose radius is at
string scale. If such a solution were found, it would be presumably
dual to ${\cal N}=2$ CS coupled to 1 adjoint matter.

\subsec{D2-branes in massive IIA theory}

Another way of obtaining Chern-Simons coupling is to consider
D2-branes in massive IIA theory, with $F^{RR}_{(0)}=k$. Fundamental
matter fields can be introduced by adding D6 or D8-branes. Let us
consider the system of $N$ D2-branes and $N_f$ D6-branes, with the
D2 lying inside the world volume of D6 in flat space. There are 3
${\cal N}=2$ adjoint matter fields $\Phi_i$, $i=1,2,3$, as well as
fundamental and anti-fundamental matter fields $Q^j, \tilde Q^j$,
$j=1,\cdots,N_f$. $\Phi_1$ corresponds to the 2 coordinates of the
D2-brane transverse to the D6, where as $\Phi_{2,3}$ correspond to
the transverse coordinates of the D2 within the world volume of D6.
There is a superpotential of the form \eqn\awuptd{ W = {\rm
Tr}\Phi_1[\Phi_2,\Phi_3] + \tilde Q^j \Phi_1 Q^j } In the IR, it is
conceivable that $Q^j,\tilde Q^j,\Phi_2,\Phi_3$ remain of dimension
$\half$, whereas $\Phi_1$ becomes a dimension 1 field, so that $W$,
of dimension 2, is marginal. The kinetic term for $\Phi_1$ then
becomes irrelevant. If we introduce a small deformation of the
superpotential $\half\epsilon{\rm Tr}\Phi^2$, so that the total
superpotential is \eqn\awuaaptd{ W = \half\epsilon {\rm Tr}\Phi_1^2
+{\rm Tr}\Phi_1[\Phi_2,\Phi_3] + \tilde Q^j \Phi_1 Q^j, } we can
then integrate out $\Phi_1$ and obtain an equivalent superpotential
\eqn\awuabbptd{ W = {1\over 2\epsilon}{\rm Tr}\left([\Phi_2,\Phi_3]
+ Q^j\tilde Q^j\right)^2 } When $\epsilon\ll k$, the superpotential
dominates the interaction due to Chern-Simons coupling, and acquires
positive anomalous dimension (i.e. positive beta function). As in
our previous discussion of ${\cal N}=3$ Chern-Simons-matter theory,
the theory with superpotential \awuabbptd\ flows to the point where
the coefficient $1/\epsilon$ becomes $\pi/k$, and the theory becomes
nothing but the ${\cal N}=3$ $U(N)$ Chern-Simons theory coupled
minimally to an adjoint hypermultiplet $(\Phi_2,\Phi_3)$ and $N_f$
fundamental hypermultiplets $(Q^j,\tilde Q^j)$.

It is thus conceivable that the world volume theory of D2-D6 in
massive IIA theory flows to the ${\cal N}=3$ Chern-Simons-matter
theory under the deformation given by a small mass term for
$\Phi_1$. Recall that the ${\cal N}=2$ Chern-Simons-matter theory
also flows to the ${\cal N}=3$ theory, but in the opposite direction
along the line of the coefficient of the superpotential. It would be
interesting to study the near horizon geometry of D2-D6 system in
massive IIA theory, and see if the above RG flow can be described in
the gravity dual.

\subsec{Hints from spin chains}

The spin chain analysis in the previous section provides several
useful hints on the holographic dual of ${\cal N}=3$ CS coupled to
one adjoint hypermultiplet. There is a tower of short
representations of $OSp(3|4)\times SU(2)_f$ generated by ${\rm
Tr}Q^J$ with spin $({J\over 2},{J\over 2})$ under $SU(2)_R\times
SU(2)_f$. This resembles the spectrum of KK modes of a 7-dimensional
supergravity on $AdS_4\times S^3$. The fact that the $SU(2)_R$ and
$SU(2)_f$ impurities have the same spectrum further reinforces the
idea that the two $SU(2)$'s should appear in the dual geometry in a
symmetric fashion, as left and right rotations of an $S^3$. The
supercoset $OSp(3|4)/SO(3,1)$ has $AdS_4\times S^3$ as its bosonic
part, and the correct symmetry group.

The ${\cal N}=2$ CS with one adjoint matter $\Phi$, deformed by the
superpotential ${\rm Tr}\Phi^4$, should be dual to a theory of
gravity in $AdS_4$, whose massless sector is ${\cal N}=2$
supergravity coupled to a universal hypermultiplet. It might be
possible that when the 't Hooft coupling is large, all other fields
become infinitely massive, and only the supergravity sector remains.

We leave the detailed analysis of the holographic dual geometries to
future investigation.

\newsec{Concluding remarks}

We found a surprisingly large class of three dimensional ${\cal
N}=2$ SCFTs with Lagrangian descriptions, whose couplings can be
made arbitrarily weak, as superpotential deformations of ${\cal
N}=2$ Chern-Simons-matter theories. This allows perturbative
understanding of the SCFTs in $1/k$. For $U(N)$ theories, we can
also study the $1/N$ expansion. We have seen evidences for the
existence of a weakly coupled string theory dual in the large $N$
limit, although we have not been able to construct the $AdS_4$ dual
in a supergravity regime. It is possible that the holographic dual
of these theories are described by string theory on $AdS_4$ whose
radius is at string scale. It would be interesting to investigate
the $\alpha'$ corrections in the brane constructions in type II
string theories that give rise to Chern-Simons couplings, as well as
exploring $AdS_4$ compactifications of massive IIA
supergravity/string theory. In the example of ${\cal N}=3$ theory,
we have seen that certain sectors of long operators can be described
by (non-integrable) spin chains. One may learn about the possible
dual string sigma model from this.

The rich structure of RG flows among the superpotential deformations
of ${\cal N}=2$ CS-matter theories is intriguing. In the context of
M-theory compactified on $AdS_4$ times a Sasakian-Einstein
seven-manifold, there is also a rich structure of holographic RG
flows, as explored in \refs{\CorradoNV, \AhnZY, \AhnAQ}. It would be
interesting to see if there are connections between them. More
ambitiously, one may wonder if there are connections of the
CS-matter theories to the vast number of $AdS_4$ string vacua in
flux compactifications (see for example \DouglasES).

\bigskip

\centerline{\bf Acknowledgement} We are grateful to N. Arkani-Hamed,
C. Beasley, S. Gukov, C. Herzog, D. Jafferis, H. Lin, J. Marsano, L.
Motl, A. Neitzke, V. Pestun, D. Robbins, J. Schwarz, D. Shih, A.
Strominger, A. Tirziu, E. Witten for useful discussions and
correspondences. DG is supported in part by DOE grant
DE-FG02-91ER40654. XY is supported by a Junior Fellowship from the
Harvard Society of Fellows.

\bigskip

\appendix{A}{The ${\cal N}=3$ Lagrangian}

In this appendix we write the Lagrangian of ${\cal N}=3$ Chern-Simons theory coupled to
$N_f$ hypermultiplets in a {\sl real} representation $R$
 in component fields with manifest $USp(2N_f)\times SU(2)_R$ symmetry.
Let $a, b, \cdots$ be $SU(2)_R$ indices, $A,B, \cdots$ indices for the fundamental representation of
$USp(2 N_f)$, and $\alpha,
\beta, \cdots$ $SO(2,1)$ spinor indices.

The components of the matter fields are scalars $q^{Aa}$ and
fermions $\psi^{Aa}_\alpha$, with reality condition \eqn\realitya{
\eqalign{ & (q^\dagger)_{Aa} = \omega_{AB} \epsilon_{ab} q^{Bb},\cr
& \bar\psi_{Aa}^\alpha = \omega_{AB} \epsilon_{ab}
\epsilon^{\alpha\beta} \psi^{Bb}_\beta. } } where $\omega_{AB}$ is a
symplectic form. They are related to the fields $(Q,\tilde Q)$ in
the ${\cal N}=2$ notation by \eqn\relaaa{ q^{A1} = (Q, \tilde
Q),~~~~~ q^{A2} = (-\bar{\tilde Q},\bar Q). } The on-shell
supersymmetry transformations are \eqn\susytrnas{\eqalign{
\delta_{ab}^{\alpha} A_{\beta \gamma}^m &= {4\pi\over k} q_{A(a} T^m
\psi^A_{b)(\gamma} \delta^{\alpha}_{\beta)}, \cr
\delta_{ab}^{\alpha} q_{Ac} &= \psi^{\alpha}_{A(a}\epsilon_{b)c},\cr
\delta_{ab}^{\alpha} \psi^{\beta}_{Ac} &= -i\nabla^{\alpha \beta}
q_{A(a}\epsilon_{b)c}+ {4\pi\over k} (q_{Bc} T^m q^B_{(a}) T^m
q_{b)A}.  }} Introducing auxiliary fields $s^m_{ab}={4\pi\over
k}q_{A(a} T^m q^A_{b)}$, $\chi^m_{ab}=-{4\pi i\over k}q_{A(a} T^m
\psi^A_{b)}$, $\chi^m=-{4\pi i\over k}q_{Aa} T^m \psi^{Aa}$, the
Lagrangian can be written as \eqn\nthelag{\eqalign{ L&={k\over4\pi}
\left[ CS(A) +{\rm Tr}(D^{ab}s_{ab}-{1\over
2}\chi^{ab}\chi_{ab}+\chi\chi +{1\over 6}s^{ab}[s_{bc},{s^c}_{a}])
\right] \cr &~~~+ {1\over 2}|\nabla_{\mu} q_{aA}|^2 + {1\over
2}q_{Aa} D^{ab} {q^A}_b - {1\over 4}|s_{ab} q^{Ac}|^2 \cr & ~~~ +
{i\over 2} \psi_{aA}\gamma^{\mu} \nabla_{\mu} \psi^{aA} -{1\over 2}
\psi_{A}^{a} s_{ab} \psi^{Ab} + i{q_A}^a\chi_{ab}\psi^{Ab} +
iq_{Aa}\chi\psi^{Ab}. } } To study BPS (chiral) operators one needs
to pick a chiral supersymmetry generator $\delta^+ = u^a u^b
\delta_{ab}$ for some twistor variable $u^a$. Then $Q_A(u)\equiv u^a
q_{aA}$ are the chiral fields.

\appendix{B}{Thermodynamics of the abelian theory at large $N_f$}

\subsec{${\cal N}=2$ theory with equally charged matter in flat
space}

In this section we will consider the free energy of the ${\cal N}=2$
theory with equally charged matter fields in flat space. Keeping the
auxiliary fields in the vector super-multiplet, we can integrate out
the matter fields, and compute the thermal partition function via
the path integral \eqn\athint{\eqalign{ & \int dAdDd\sigma d\chi
d\bar\chi \exp\left[-{N_f\over \lambda}\int (\omega_{CS}(A) +
2D\sigma-\bar\chi\chi)\right. \cr & \left. - N_f {\rm Str} \ln
\left( \eqalign{-D^\mu D_\mu + \sigma^2+D &~~~~~~~~ \bar\chi
\cr\chi~~~~~~~~~~~~ & ~~~~i{\slash\!\!\!\!}D +\sigma }
 \right) \right] }}
on ${\bf R}^2\times S^1_\beta$. In the large $N_f$ limit, the path
integral by the saddle point contribution. We must be careful with
the definition of this path integral: the superspace action \suppac\
is linear in the auxiliary field $D$, and integration over $D$
results in a delta functional that gives rise to a (Euclidean)
action which is bounded from below. While keeping $D$ in \athint, we
cannot naively minimize the action with respect to $D$, as it would
obviously give a divergent answer. The correct prescription is to
extremize the effective action with respect to $D$ first, and then
minimize the effective action with respect to the other
fields.\foot{To see this more explicitly, consider a potential
function $V(D,x)=xD+f(x)$. ``Integrating out" $D$ sets the potential
to $f(0)$. On the other hand, ``integrating out" $x$ amounts to a
Legendre transform, $D=-f'(x)$. The potential $V(D,x(D))$ is
extremized with respect to $D$ at $x(D)=0$. Whether it is a local
maximum or minimum in $D$ depends on the sign of $f''(0)$. }

The saddle points involve translational invariant (as well as
rotational invariant in the case of finite temperature) field
configurations. So we can set $A_i=0, \chi=0$, and $D,\sigma,
A_0=\alpha$ to constants. We then have the very simple effective
action in the constant modes $D,\sigma,\alpha$, at finite
temperature $T=1/\beta$, \eqn\sfee{ {S_{eff}} = N_f A\beta \left[
-{2D\sigma\over\lambda} + \beta^{-1}\sum_{n\in{\bf Z}}\int
{d^2p\over (2\pi)^2} \ln {p^2+({2\pi
\over\beta}n+\alpha)^2+\sigma^2+D\over
p^2+({2\pi\over\beta}(n+\half)+\alpha)^2 + \sigma^2} \right] }

We can dimensionally regularize the integral over spatial momenta,
and then regularize the sum over the momentum along Euclidean time
direction using zeta function regularization. The $1/\epsilon$
divergence cancel after the zeta function regularization, and we end
up with \eqn\asff{\eqalign{ S_{eff} &=N_fA\beta \left\{
-{2D\sigma\over\lambda} - {1\over 4\pi\beta}\sum_{n\in{\bf Z}}
\left[({2\pi n\over \beta}+\alpha)^2+\sigma^2+D\right] \ln
\left[({2\pi n\over \beta}+\alpha)^2+\sigma^2+D\right] \right. \cr
&\left.~~~+{1\over 4\pi\beta} \sum_{n\in{\bf Z}} \left[({2\pi
(n+\half)\over \beta}+\alpha)^2+\sigma^2\right] \ln \left[({2\pi
(n+\half)\over \beta}+\alpha)^2+\sigma^2\right] \right\} }} To
evaluate it, we shall use the formula (free energy of a harmonic
oscillator) \eqn\asnrtt{\eqalign{& \sum_{n\in {\bf Z}} \ln
\left[x^2+({2\pi n\over \beta}+\alpha)^2\right] =\beta x + \ln
(1-e^{-\beta x+i\beta \alpha}) + \ln(1-e^{-\beta x-i\beta\alpha}) }}
Integrating this, we obtain \eqn\asddin{ \eqalign{ & \sum_{n\in {\bf
Z}} \left[x^2+({2\pi n\over \beta}+\alpha)^2\right] \ln
\left[x^2+({2\pi n\over \beta}+\alpha)^2\right] \cr &={2\beta\over
3}x^3 + 4\sum_{k=1}^\infty {1+k\beta x\over \beta^2k^3} e^{-k\beta
x}\cos(k\beta\alpha)= {2\beta\over 3}x^3 +4\beta^{-2} {\cal I}(\beta
x,\beta \alpha) \cr &= {2\beta\over 3}x^3 + {2x \over
\beta}\left[{\rm Li}_2(e^{-\beta (x+i\alpha)})+{\rm Li}_2(e^{-\beta
(x-i\alpha)})\right] + {2\over \beta^2}\left[{\rm Li}_3(e^{-\beta
(x+i\alpha)})+{\rm Li}_3(e^{-\beta (x-i\alpha)})\right]  } } where
we defined the function \eqn\iit{ {\cal I}(x,\alpha) =
\sum_{k=1}^\infty {1+kx\over k^3}e^{-kx}\cos(k\alpha) =
\sum_{k=1}^\infty {\cos(k\alpha)\over k^3} + x^2 \ln \left| 2\sin
({\alpha\over 2}) \right| -{x^3\over 6}+{x^4\over 32
\sin^2({\alpha\over 2})} +{\cal O}(x^6) }
 Note
that when $x=0, \alpha=0$, \asddin\ reproduces the contribution to
the free energy from a massless free field. The effective action is
then \eqn\sfff{ S_{eff} = -N_f A\beta \left\{ {2D\sigma\over
\lambda} + {1\over 6\pi}\left[ (\sigma^2+D)^{3\over 2}-|\sigma|^3
\right] + {{\cal I}(\beta({\sigma^2+D})^{1\over 2},\beta\alpha) -
{\cal I}(\beta|\sigma|,\beta\alpha+\pi) \over \pi\beta^3} \right\} }
To separate the temperature dependence, we can rescale the variables
$\tilde\sigma=\beta\sigma$, $\tilde D=\beta^2 D$, $\tilde
\alpha=\beta\alpha$, and write \eqn\sfff{\eqalign{ S_{eff} &= -N_f
A\beta^{-2} \left\{ {2\tilde D\tilde \sigma\over \lambda} + {1\over
6\pi}\left[ (\tilde \sigma^2+\tilde D)^{3\over 2}-|\tilde \sigma|^3
\right] + {{\cal I}(({\tilde \sigma^2+\tilde D})^{1\over 2},\tilde
\alpha) - {\cal I}(|\tilde \sigma|,\tilde \alpha+\pi) \over \pi}
\right\}  }}
 We
shall first minimize the effective action with respect to $\alpha$.
This is achieved simply at $\alpha=0$. Define
$\tilde\rho=(\tilde\sigma^2+\tilde D)^{1\over 2}$. The saddle point
equations for $\tilde\rho$ and $\tilde \sigma$ are now simply
\eqn\saafst{\eqalign{& -{4\pi\over \lambda}\tilde \sigma_* = \ln
(2\sinh{\tilde \rho_*\over 2}),\cr & {2\pi\over \lambda}(\tilde
\rho_*^2-3\tilde \sigma_*^2) = \tilde \sigma_*
\ln(2\cosh{\tilde\sigma_*\over 2}). }} For $\lambda\ll 1$, this
becomes \eqn\asfff{ \tilde \sigma_* = -{\lambda\over 4\pi}\ln
(\sqrt{3}\tilde \sigma_*)\sim {\lambda|\ln\lambda|\over
4\pi},~~~~\tilde\rho_*= \sqrt{3}\tilde\sigma_* \sim
{\sqrt{3}\lambda|\ln\lambda|\over 4\pi}. } On the other hand, in the
strong coupling limit $\lambda\to\infty$, the solution is \eqn\saff{
\tilde\rho_*=\ln{3+\sqrt{5}\over 2},~~~~\tilde\sigma_* = {2\pi
\tilde\rho^2_*\over \lambda\ln2}\to0. }
 The free energy is given by
\eqn\sffex{ F(\lambda, T) = {N_fT^2\over \pi}\left\{
{2\pi(\tilde\rho_*^2-\tilde\sigma_*^2)\tilde\sigma_*\over
\lambda}+{\tilde\rho_*^3-\tilde\sigma_*^3\over 6} +
\sum_{k=1}^\infty
{(1+k\tilde\rho_*)e^{-k\tilde\rho_*}-(-)^k(1+k\tilde\sigma_*)e^{-k\tilde\sigma_*}\over
k^3} \right\} } The free energy is analytic in $\lambda$ except at
$\lambda=0$, and decreases monotonously as $\lambda$ increases. We
have \eqn\asfreae{ \eqalign{ & F(\lambda=0,T) = N_f
T^2{7\zeta(3)\over 4\pi} \simeq 0.669597 N_f T^2, \cr &
F(\lambda=\infty, T) = {N_fT^2\over \pi}\left[{3\over 4}\zeta(3) +
{1\over 6} (\ln{3+\sqrt{5}\over 2})^3 + \ln ({3+\sqrt{5}\over 2})
{\rm Li}_2({2\over 3+\sqrt{5}}) + {\rm Li}_3({2\over 3+\sqrt{5}})
\right]\cr &~~~~~~~~\simeq 0.593071 N_fT^2. } } At weak coupling, we
can expand the free energy to first subleading order as
\eqn\smallcofr{ F(\lambda,T) = N_f T^2 \left[ {7\zeta(3)\over 4\pi}
+ {\lambda^2 (\ln\lambda)^3\over 32\pi^2} +\cdots \right], ~~~~~~~~
\lambda\ll 1. }
 There exists another set of
saddle point solutions at small $\lambda$, which have nonzero
$\alpha$. Their contribution to the free energy is always smaller
than the saddle points described above. Note that the strong
coupling limit result of the free energy in \asfreae, apart from the
contribution from free fermion, is the same as that of the IR fixed
point of three dimensional $U(N_f)$ model in the infinite $N_f$
limit.

\subsec{On the sphere}

The free energy of ${\cal N}=2$ Chern-Simons-matter theory on a
sphere of radius $R=1$ is given in the large $N_f$ by the
extremizing the function \eqn\faaa{ -S_{eff}(\sigma,D,\alpha)=N_f
\left[ {8\pi\beta \over\lambda} \tilde D\tilde \sigma -
Z_B({\beta},\sqrt{\tilde \sigma^2+\tilde D},\alpha)
+Z_F({\beta},|\tilde \sigma|,\alpha)\right] } where we have defined
$D=\beta^2 \tilde D$, $\sigma = \beta \tilde\sigma$. $Z_B$ and $Z_F$
are given by \eqn\affaa{\eqalign{ Z_B(\beta,x,\alpha) & =
\sum_{l=0}^\infty (2l+1) \left[ \beta\sqrt{x^2+(l+\half)^2} +\ln
\left(1-e^{-\beta\sqrt{x^2+(l+\half)^2}+i\alpha}\right) \right.\cr
&~~~~~~~~~\left.+\ln
\left(1-e^{-\beta\sqrt{x^2+(l+\half)^2}-i\alpha}\right) \right], \cr
 Z_F(\beta,x,\alpha) &= \sum_{l=1}^\infty 2l \left[ \beta\sqrt{x^2+l^2}
+\ln \left(1+e^{-\beta\sqrt{x^2+l^2}+i\alpha}\right) +\ln
\left(1+e^{-\beta\sqrt{x^2+l^2}-i\alpha}\right) \right]. }} After
regularizing the divergent parts of the sums, we have \eqn\asffreg{
\eqalign{ Z &_B(\beta, x,\alpha)= \beta\sum_{l=0}^\infty (2l+1)
\left(\sqrt{x^2+(l+\half)^2}-(l+\half)-{x^2\over 2l+1}\right)+ {\cal
J}_B(\beta, x, \alpha),\cr
&Z_F(\beta,x,\alpha)=\beta\left[-{x^2\over 2} +\sum_{l=1}^\infty 2l
\left(\sqrt{x^2+l^2}-l-{x^2\over 2l}\right)\right] + {\cal
J}_F(\beta, x, \alpha), } } with \eqn\asjff{\eqalign{ & {\cal
J}_B(\beta, x, \alpha) = \sum_{l=0}^\infty (2l+1) \left[\ln
\left(1-e^{-\beta\sqrt{x^2+(l+\half)^2}+i\alpha}\right) + \ln
\left(1-e^{-\beta\sqrt{x^2+(l+\half)^2}-i\alpha}\right)\right],\cr &
{\cal J}_B(\beta, x, \alpha) = \sum_{l=0}^\infty 2l\left[\ln
\left(1+e^{-\beta\sqrt{x^2+l^2}+i\alpha}\right) +\ln \left(1
+e^{-\beta\sqrt{x^2+l^2}-i\alpha}\right)\right].  } } Note that the
infinite sums in \asffreg\ are convergent. ${\cal J}_B$ and ${\cal
J}_F$ are suppressed by powers of $e^{-\beta}$ in the low
temperature limit. The free energy is now analytic in $\lambda$ at
$\lambda=0$, as there is no infrared divergence.

We will now compute the free energy in the strict large $N_f$ limit
at low temperature. At the saddle point $\alpha=0$. To leading order
in $e^{-\beta}$, the saddle point equations for $\tilde\sigma$ and
$\tilde \rho=\sqrt{\tilde \sigma^2+\tilde D}$ are \eqn\saddeee{
\eqalign{ & {8\pi\over \lambda}\tilde\sigma_* -
2e^{-\beta/2}+{\pi^2\over 4}\tilde\rho_*^2+{\cal O}(e^{-\beta})=0,
\cr & {8\pi\over \lambda} (\tilde\rho_*^2-\tilde\sigma_*^2) - \tilde
\sigma_* + {\cal O}(e^{-\beta})=0 } } The solutions are
\eqn\addsol{\eqalign{& \tilde \sigma_*={\lambda\over 4\pi
(1+{\lambda^2\over 256})}e^{-\beta/2}, ~~~~~~ \tilde\rho_*
={\lambda\over 4\sqrt{2}\pi\sqrt{1+{\lambda^2\over 256}}}
e^{-\beta/4}. } } The free energy is then given by \eqn\freelowt{ F
= N_f \left[2 e^{-\beta/2} + (5-{\beta\over 32\pi^2}{\lambda^2\over
1+{\lambda^2\over 256}}) e^{-\beta}+{\cal O}(e^{-3\beta/2})\right] }
This expression is valid for all values of $\lambda$ in the low
temperature limit. The term of order $\beta e^{-\beta}$ in
\freelowt\ can be reproduced by summing up diagrams as in Figure 7.
When higher order terms in $e^{-\beta}$ are included, diagrams
involving the $|\phi|^6$ vertices also contribute.

\vskip 0.5cm \centerline{\vbox{\centerline{
\hbox{\vbox{\offinterlineskip \halign{&#&\strut\hskip0.2cm \hfill
#\hfill\hskip0.2cm\cr \epsfysize=.8in \epsfbox{chain.1}  \cr }}}}
{{\bf Figure 7.} The solid lines and double lines represent the
scalar and fermion respectively.}}} \vskip 0.5cm

Let us note that the saddle point approximation in the strict large
$N_f$ limit is not sufficient to extract the spectrum of low
dimensional operators. In the free theory for instance, the lowest
dimensional gauge invariant operator other than 1 is $\bar \phi^i
\phi^j$, of dimension 1. There are $N_f^2$ such operators. They
contribute to the partition function $N_f^2 e^{-\beta}$. In order to
see this in the low temperature expansion of the partition function,
we would need $N_f^2 e^{-\beta} \ll 1$. The saddle point
approximation breaks down in this regime. For example, we must
integrate out $\alpha$ in the full path integral in order to see
that the partition function only sums up gauge invariant states.
This becomes invisible in the saddle point approximation, where
$\alpha$ is set to zero. In fact, a simple diagrammatics reveals
that the anomalous dimensions of finite dimensional operators are
subleading in $1/N_f$, and vanish in the infinite $N_f$ limit.

\subsec{${\cal N}=2$ theory with oppositely charged matter}

The effective action for the saddle point is of the form \eqn\efeat{
-{\beta^2\over N_f A}S_{eff}=
{2D\sigma\over\lambda}+{f_B(\sqrt{\sigma^2+D})
+f_B(\sqrt{\sigma^2-D})-2f_F(|\sigma|)\over\pi} } where $f_B,f_F$
are given by \eqn\assaff{\eqalign{ & f_B(x) = {\cal
I}(x,0)+{x^3\over 6} = \sum_{k=1}^\infty {1+kx\over
k^3}e^{-kx}+{x^3\over 6},\cr & f_F(x) = {\cal I}(x,\pi)+{x^3\over 6}
= \sum_{k=1}^\infty (-)^k{1+kx\over k^3}e^{-kx}+{x^3\over 6}.
 }}
as before. A subtlety is that, the saddle point of interest may lie
in the regime $|D|>\sigma^2$, and we would need to consider $f_B(x)$
with imaginery $x$. This is not a serious problem. In fact, one can
analytically continue $f_B(x)$, so that $f_B(x)$ is real for both
real and imaginery $x$, and is an even function in $x$. The saddle
point equations are now written as \eqn\sadpoteqna{\eqalign{ &
-{4\pi\over \lambda}\sigma = \ln \left|{\sinh{\sqrt{\sigma^2+D}\over
2}\over\sinh{\sqrt{\sigma^2-D}\over 2}}\right| , \cr & {2\pi\over
\lambda}D = \sigma \ln\left|{\cosh^2({\sigma\over 2}) \over
\sinh{\sqrt{\sigma^2+D}\over 2}\sinh{\sqrt{\sigma^2-D}\over
2}}\right|. }} The only saddle point is $D=\sigma=0$, and hence the
free energy is identical to that of the free theory. For the ${\cal
N}=3$ theory, the only difference in the effective action \efeat\ is
that one replaces $\sigma$ and $D$ with $su(2)$-valued $\sigma_a
\gamma^a$ and $D_a\gamma^a$, where $\gamma^a$ are Pauli matrices,
and take the trace over the doublet. The saddle points lead to the
same result, i.e. the free energy of the ${\cal N}=3$ theory is the
same as that of the free theory to leading order in $1/N_f$.

\appendix{C}{Thermodynamics of the free $U(N)$ theory on the sphere}

In this appendix, we study the operator spectrum of the free theory
($\lambda=0$) from the point of view of thermodynamics. We are
essentially reproducing here some of the results of \SchnitzerQT.
One might naively expect the thermodynamics to be trivial for the
free theory. This would be the case in flat space, but is not the
case on the sphere due to the restriction to gauge invariant states
\AharonySX. The partition function of the free theory reduces to a
unitary matrix model of the form \eqn\untt{ \int [dU]_{U(N)}
e^{-S_{eff}(U)} } where $U$ is the holonomy of the gauge field along
the thermal circle, $U=e^{i\alpha}$, $\alpha$ being the zero mode of
$A_0$. The matrix model action is given by \eqn\effctaa{ S_{eff}(U)
= -2N_f \sum_{n=1}^\infty {z_S(x^n)+(-)^{n+1}z_F(x^n)\over n} ({\rm
tr} U^n + {\rm tr} U^{-n}), } where $x=e^{-\beta}$, $z_S(x)$ and
$z_F(x)$ are the partition functions of a conformally coupled scalar
particle and spin $1/2$ particle on the sphere, \eqn\aqcsf{
\eqalign{ & z_S(x) = \sum_{l=0}^\infty (2l+1)x^{l+\half} = {x^\half
(1+x)\over (1-x)^2}, \cr & z_F(x) = \sum_{l=1}^\infty 2l x^l =
{2x\over (1-x)^2}. } } Diagonalizing the unitary matrix $U={\rm
diag} (e^{i\alpha_1},\cdots,e^{i\alpha_N})$, the matrix integral
becomes \eqn\itntmatt{\eqalign{ &\int \prod d\alpha_i \exp\left\{ -
\left[ \sum_{i\not=j} \sum_{n=1}^\infty
{\cos(n(\alpha_i-\alpha_j))\over
n}-4N_f\sum_i\sum_{n=1}^\infty{z_S(x^n)+(-)^{n+1}z_F(x^n)\over
n}\cos(n\alpha_i) \right] \right\} \cr &= \int \prod d\alpha_i
\exp\left\{ - \left[ \sum_{i\not=j} \sum_{n=1}^\infty
{\cos(n(\alpha_i-\alpha_j))\over
n}-2N_f\sum_i\sum_{n=1}^\infty{\cosh({n\beta\over 2})+(-)^{n+1}\over
n \sinh^2({n\beta\over 2})}\cos(n\alpha_i) \right] \right\} } } In
the large $N$ limit, we can represent the eigenvalues by the
eigenvalue density $\rho(\theta)$, with the property
\eqn\astdens{\rho(\theta)\geq 0,~~~~~\int_0^{2\pi} d\theta
\rho(\theta)=1. } The potential function in \itntmatt\ now becomes
\eqn\aintpo{ N^2 \sum_{n=1}^\infty {1\over n}\left[ \rho_n^2 -
2c\rho_n {\cosh({n\beta\over 2})+(-)^{n+1}\over \sinh^2({n\beta\over
2})} \right] } where $\rho_n=\int d\theta \rho(\theta)
\cos(n\theta)$, and $c=N_f/N$. Defining the function \eqn\afuncsa{
f_n(\beta) = {\cosh({n\beta\over 2})+(-)^{n+1}\over
\sinh^2({n\beta\over 2})}, } we can rewrite \aintpo\ as \eqn\abtss{
N^2 \sum_{n=1}^\infty {(\rho_n-cf_n(\beta))^2-c^2f_n(\beta)^2\over
n} } At low temperatures, the saddle point is given by $\rho_n =
cf_n(\beta)$, and the free energy of the theory on the sphere is
\eqn\ffte{F_{low}(\beta) = N_f^2\sum_{n=1}^\infty {f_n(\beta)^2\over
n} } \ffte\ ceases to be valid when $f_1(\beta)\sim c^{-1}$. At high
temperatures, the saddle points are very different since \astdens\
severely constrains the $\rho_n$'s. The free energy in the high
temperature limit is simply given by the flat space result
\eqn\sitne{ F_{high}(\beta) \simeq NN_f{7\zeta(3)\over\beta^2} }
There is a transition from $N_f^2$ degrees of freedom at low
temperature to $NN_f$ degrees of freedom at high temperature.

At nonzero coupling, we can still integrate out all the matter
fields while keeping the Chern-Simons auxiliary fields, and obtain a
path integral of the form \eqn\btbint{\eqalign{ & \int dAdDd\sigma
d\chi d\bar\chi \exp\left[-{N\over \lambda}\int (\omega_{CS}(A) +
2D\sigma-\bar\chi\chi)\right. \cr & \left. - N_f {\rm Str} \ln
\left( \eqalign{-D^\mu D_\mu + \sigma^2+D &~~~~~~~~ \bar\chi
\cr\chi~~~~~~~~~~~~ & ~~~~i{\slash\!\!\!\!}D +\sigma }
 \right) \right] }} where the ``${\rm Str}$" in the second term in the
 effective action involves a trace in the fundamental representation of
 $U(N)$.
However, we can no longer integrate out all but one field in the
Chern-Simons multiplet. The saddle point approximation to the path
integral \btbint\ at large $N$ is no longer valid, since there are
$\sim N^2$ fields in \btbint\ while the action is only multiplied by
$N$.

\appendix{D}{Two-loop anomalous dimensions}

\subsec{The two-loop anomalous dimension of $\bar\phi \phi$}

It was predicted that in ${\cal N}=2$ CS-matter theory, the
anomalous dimension of $\bar\phi^i \phi_j$ is zero, since it lies in
the same supermultiplet as the $U(N_f)$ flavor current. It is
nevertheless instructive to verify this through an explicit two-loop
computation. The contributions come from wave function
renormalization, as shown in Fig 8, as well as 1PI diagrams in Fig
9. We will work in Feynman gauge, regularizing the loop integrals
with dimensional reduction method.

\vskip 0.3cm \centerline{\vbox{ \hbox{
\centerline{\vbox{\offinterlineskip \halign{&#&\strut\hskip0cm
\hfill #\hfill\hskip0cm\cr & \epsfysize=0.6in \epsfbox{phiphi.1}
 &~~~~& \epsfysize=0.6in \epsfbox{phiphi.2}
 &~~~~& \epsfysize=0.6in \epsfbox{phiphi.3}
 &~~~~& \epsfysize=0.6in \epsfbox{phiphi.4}
  &~~~~& \epsfysize=0.6in \epsfbox{phiphi.5} & \cr  &
  $(a)$ && $(b)$ && (c) && (d) && (e) & \cr
}}}} { {\bf Figure 8:} The solid and double lines stand for scalar
and fermion propagators, respectively. The shaded bubble in $(b)$
represents matter loops. The 1-loop gauge and ghost bubbles cancel.
}}} \vskip 0.3cm

For simplicity we will do the calculation in the planar limit (for
fundamental and adjoint matter), although the diagrams in Fig 8 and
Fig 9 are not drawn as planar diagrams. The circle in Fig 9
represents the operator insertion of $\bar\phi^i \phi_j$. We have
only shown diagrams with nonzero momentum integral. One must be
careful with the planar combinatorics. This is most easily taken
care of by keeping the propagators of auxiliary fields $\sigma,
D,\chi$, which have been integrated out in Fig 8 and 9.

\vskip 0.3cm \centerline{\vbox{ \hbox{
\centerline{\vbox{\offinterlineskip \halign{&#&\strut\hskip0cm
\hfill #\hfill\hskip0cm\cr & \epsfysize=0.6in \epsfbox{phiphiop.1}
 &~~~~& \epsfysize=0.6in \epsfbox{phiphiop.2} & \cr  &
  $(f)$ && $(g)$ & \cr
}}}} \centerline{ {\bf Figure 9.} }}} \vskip 0.2cm

Some relevant loop integrals are \eqn\loapta{ \eqalign{ &\int
{d^3k\over (2\pi)^3}{d^3l \over (2\pi)^3}{1\over k^2
l^2(k+l)^2}={\ln\Lambda\over 16\pi^2} \cr  &  \int {d^3k\over
(2\pi)^3}{d^3l \over (2\pi)^3}{2k\cdot l\over k^2 l^2(p+k+l)^2} =
{p^2\over 3}{\ln\Lambda\over 16\pi^2}+\cdots\cr & \int {d^3k\over
(2\pi)^3}{d^3l \over (2\pi)^3} {(\epsilon_{\mu\nu\rho}p^\mu k^\nu
l^\rho)^2\over k^2 l^2 (k+l)^2 (p+k)^2(p+k+l)^2}={p^2\over
6}{\ln\Lambda\over 16\pi^2}+\cdots } } where we omitted power
divergent and finite terms. The wave function renormalization is
computed to be \eqn\waenfa{\eqalign{ \delta Z &= ({2\pi\over
k})^2(-{5\over 3}NN_f+{8\over 3}NN_f+{1\over 3}N^2+0-{4\over 3}N^2)
{\ln\Lambda\over 16\pi^2}\cr &= ({2\pi\over k})^2(NN_f-N^2)
{\ln\Lambda\over 16\pi^2}, } } where the terms in the parenthesis in
the first line come from diagrams $(a-e)$. The contributions from
the diagrams $(f)$ and $(g)$ to the anomalous dimension precisely
cancel the correction due to $\delta Z$, confirming that $\bar\phi^i
\phi_j$ has zero anomalous dimension at two-loop.

For the ${\cal N}=2$ theory with $M$ adjoint matters (in the planar
limit), the wave function renormalization is \eqn\waenfadj{\eqalign{
\delta Z &= ({4\pi\over k})^2\left[-{1\over 3}N^2(5M+2)+{8\over
3}N^2M+{1\over 2}N^2+{4\over 3}N^2-{2\over 3}N^2\right]
{\ln\Lambda\over 16\pi^2}\cr &= ({4\pi\over k})^2\left(N^2M+{1\over
2}N^2\right) {\ln\Lambda\over 16\pi^2}, } } where the five terms in
the first line come from diagrams $(a-e)$. The contribution from
$(f)$, $(g)$ is \eqn\apfg{ ({4\pi\over k})^2 \left[
-N^2(M+2)+{3\over 2}N^2\right] {\ln\Lambda\over 16\pi^2} } This
precisely cancels \waenfadj, confirming that the anomalous dimension
of ${\rm Tr}(\bar\Phi^i\Phi^j)$ is zero. On the other hand, the
operator ${\rm Tr}(\Phi^i \Phi^j)$ is a chiral primary, whose
anomalous dimension is entirely due to the renormalization of the
$U(1)_R$ charge of $\Phi$. For this operator, the contribution from
$(f)$ and $(g)$ is \eqn\apfbxg{ ({4\pi\over k})^2 \left(
N^2M+{3\over 2}N^2\right) {\ln\Lambda\over 16\pi^2} } We find that
the anomalous dimension of ${\rm Tr}\Phi^a \Phi^b$ is \eqn\anroaa{
\Delta-1=-2(M+1)\lambda^2, } or equivalently, $q^R_\Phi={1\over
2}-{(M+1)}\lambda^2$, at two-loop. When $M$ is even, this result can
be reproduced by simply comparing to the ${\cal N}=3$ theory with
$M/2$ adjoint hypermultiplets, as in described section 2.

\subsec{Computation of the two-loop anomalous dimension of twist-1
operators}

In this section we compute the two-loop diagrams of Figure 4, for the anomalous dimension
of twist-1 operators at large spin $n$.

The loop integral from diagram $4(a)$ is evaluated as follows:
\eqn\dgmaa{\eqalign{& {1\over 4}\sum_{i=0}^{n-2}\sum_{j=0}^{n-2-i}
\int {d^3k\over (2\pi)^3} {d^3l\over (2\pi)^3}
{\epsilon_{\mu\nu\alpha}k^\alpha \epsilon_{\rho\nu\beta}l^\beta
\Delta^\mu \Delta^\rho\over k^2 l^2 (k+l+p)^2}
(\Delta\cdot(k+l+p))^i (\Delta\cdot (l+p))^j (\Delta\cdot
p)^{n-2-i-j}\cr & = {1\over 4}\sum_{i,j} \int {d^3k\over (2\pi)^3}
{d^3l\over (2\pi)^3} {((k-l)\cdot \Delta) ((l-p)\cdot \Delta)\over
k^2 (k-l)^2 (l-p)^2 } (\Delta\cdot k)^i (\Delta \cdot l)^j
(\Delta\cdot p)^{n-2-i-j}\cr &=\sum_{i,j} {2 \pi^3\over
4(2\pi)^6}\ln\Lambda \int_0^1 dx \int_0^{1-x} dy {(\Delta\cdot
p)^n\over (1-y)^{3/2} (x+y-{x^2\over 1-y})^{3/2}} ({x\over
1-y}-1)({x\over 1-y})^i\cr &~~~ \times ({y\over x+y-{x^2\over
1-y}}-1) ({y\over x+y-{x^2\over 1-y}})^{i+j+1} \cr &=\sum_{i,j}
{2\pi^3\over 4(2\pi)^6} (\Delta\cdot p)^n\ln\Lambda \int_0^1
dx\int_0^{1-x} dy {uv\over (xy)^{3\over 2}} (1-u)^{i+{3\over 2}}
(1-v)^{i+j+{3\over 2}}, } } where we have defined $u\equiv x/(1-y)$,
$v\equiv y/(x+y-{x^2\over 1-y})$, and have thrown away power
divergences. In the $n\to\infty$ limit, the integral is in fact
finite, and is given by \eqn\dgmssa{ {1\over 64\pi^2}(\Delta\cdot
p)^n\ln\Lambda }

The contribution from (b) is \eqn\dgmbb{ \eqalign{ & {1\over
4}\sum_{i,j} \int {d^3k\over (2\pi)^3} {d^3l\over (2\pi)^3}
{\epsilon_{\rho\mu\alpha} (l-k)^\alpha \epsilon_{\sigma\nu\beta}
(p-l)^\beta \epsilon^{\rho\sigma\eta} \epsilon_{\tau\eta\gamma}
(p-k)^\gamma \Delta^\mu \Delta^\nu (p+k)^\tau\over k^2 (k-l)^2
(p-l)^2 (p-k)^2} \cr &~~~~~~~~~~~~~~\times( \Delta\cdot k)^i
(\Delta\cdot l)^j (\Delta\cdot p)^{n-2-i-j} }} Let us examine the
integral over $l$, \eqn\intlll{\eqalign{ &\int d^3l {(l-k)^\alpha
(p-l)^\beta (\Delta\cdot l)^j\over (k-l)^2(p-l)^2}\cr &= -\int_0^1
d^3l{(l+x(p-k))^\alpha (l+(1-x) (k-p))^\beta (\Delta\cdot
(l+xp+(1-x)k))^j\over (l^2+x(1-x)(k-p)^2)^2} \cr &=
(\cdots)\delta^{\alpha\beta} + \Delta^\alpha(\cdots)^\beta +
\Delta^\beta (\cdots)^\alpha }} When contracted with
$\epsilon^{\rho\sigma\eta}\epsilon_{\rho\mu\alpha}\epsilon_{\sigma\nu\beta}\Delta^\mu
\Delta^\nu$, this is just zero. Hence the contribution from diagram
(b) vanishes.

Diagram (c) is given by \eqn\dgmcc{\eqalign{ & {1\over
4}\sum_{i=0}^{n-1} \int {d^3k\over (2\pi)^3} {d^3l\over (2\pi)^3}
{\epsilon_{\mu\nu\alpha}(k+l)^\alpha \epsilon_{\rho\sigma\beta}
l^\beta \epsilon_{\eta\tau\gamma}k^\gamma \epsilon^{\rho\nu\eta}
(2p+k)^\tau (2p+2k+l)^\sigma \Delta^\mu\over k^2 l^2 (k+l)^2 (p+k)^2
(p+k+l)^2 }\cr &~~~~~~~~~~~~~~~\times ((p+k+l)\cdot \Delta)^i
(p\cdot \Delta)^{n-1-i}\cr &= \sum_{i=0}^{n-1} \int {d^3k\over
(2\pi)^3} {d^3l\over (2\pi)^3} {\epsilon_{\mu\nu\alpha}
\epsilon_{\rho\sigma\beta} \epsilon_{\eta\tau\gamma}
\epsilon^{\rho\nu\eta}  l^\alpha (l-k)^\beta k^\gamma (p+k)^\sigma
p^\tau \Delta^\mu\over k^2 l^2 (l-k)^2 (p+k)^2 (p+l)^2 }((p+l)\cdot
\Delta)^i (p\cdot \Delta)^{n-1-i} }} We will do the integral in two
steps. First, the integral over $k$ \eqn\intkkks{ \eqalign{ &\int
d^3k {(l-k)^\beta k^\gamma (p+k)^\sigma \over k^2(k-l)^2(k+p)^2} \cr
&= 2\int_0^1 dx\int_0^{1-x} dy {(-k+(1-x)l+yp)^\beta
(k+xl-yp)^\gamma (k+xl+(1-y)p)^\sigma \over \left[
k^2+x(1-x)l^2+y(1-y)p^2+2xyp\cdot l\right]^3} }} When multiplied by
$\epsilon_{\mu\nu\alpha} \epsilon_{\rho\sigma\beta}
\epsilon_{\eta\tau\gamma} \epsilon^{\rho\nu\eta} l^\alpha p^\tau$,
the integral \intkkks\ simplies drastically to \eqn\sintkk{
{3\pi^2\over 2} \int_0^1 dx\int_0^{1-x} dy {-x\delta^{\beta\gamma}
l^\sigma + (1-x) \delta^{\gamma\sigma} l^\beta\over
\sqrt{x(1-x)l^2+y(1-y)p^2+2xyp\cdot l}} } The full integral \dgmcc\
is now \eqn\sgmccc{ \eqalign{ & {3\pi^2\over
(2\pi)^6}{\Gamma({5\over 2})\over \Gamma({1\over
2})}\sum_{i=0}^{n-1} \int_0^1 dx\int_0^{1-x} dy \int_0^1 dz
\int_0^{1-z} dw (zx(1-x))^{-{1\over 2}} \cr &~~~\times \int d^3l
{(p\cdot\Delta)^{n-1-i}((p+l)\cdot\Delta)^i \left[-(l\cdot\Delta)
(p\cdot l)+(p\cdot\Delta) l^2\right] \over \left[l^2+2(w+{zy\over
1-x}) p\cdot l + (w+{zy(1-y)\over x(1-x)})p^2\right]^{5\over 2}} \cr
&\longrightarrow {3\pi^3\over (2\pi)^6} (p\cdot\Delta)^n\ln\Lambda
\sum_{i=0}^{n-1}\int_0^1 dx\int_0^{1-x} dy \int_0^1 dz \int_0^{1-z}
dw (zx(1-x))^{-{1\over 2}} \cr &~~~~\times\left[2+(i-2)(w+{yz\over
1-x}) \right] (1-w-{yz\over 1-x})^{i-1} } } This integral is also
finite in the $n\to\infty$ limit, and is given by \eqn\cddccc{ {9\ln
2\over 16\pi^2}(p\cdot\Delta)^n\ln\Lambda }

Diagram (d) involves the 1-loop correction to the gauge field
propagator from the loop of gauge fields, ghosts, as well as the
matter fields. The contributions from the gauge field loop and the
ghost loop cancel. The former is given by \eqn\gaugdprop{ \eqalign{
& {1\over 2}\int d^3l
{\epsilon_{\rho\mu\eta}\epsilon_{\nu\sigma\tau}
\epsilon^{\sigma\rho\alpha}l_\alpha
\epsilon^{\eta\tau\beta}(k+l)_\beta\over l^2(k+l)^2} = {1\over
2}\int d^3l {l_\mu (k+l)_\nu+l_\nu (k+l)_\mu\over l^2(k+l)^2}, }}
where the latter is \eqn\ghosttp{ -\int d^3l{(k+l)_\mu l_\nu \over
l^2 (k+l)^2}  } Indeed they cancel as expected from the pure
Chern-Simons theory \AlvarezGaumeWK. So far we have computed the
2-loop diagrams involving gauge interactions only, and found that
the anomalous dimension of the twist-1 operator
$J_{\mu_1\cdots\mu_n}$ to order $\lambda^2$ is bounded in the large
spin $n$ limit.

Now we consider the correction of order $\lambda^2 N_f/N$, coming
from the matter loop in diagram (d). The correction to the gauge
field propagator is \eqn\matterfipop{ \eqalign{ &\int {d^3p\over
(2\pi)^3} { (2p+k)^\mu (2p+k)^\nu - {\rm Tr}\left[ \gamma^\mu
{\slash \!\!\!}p \gamma^\nu ({\slash \!\!\!}p+{\slash \!\!\!}k)
\right] \over p^2(p+k)^2} \cr & = \int {d^3p\over (2\pi)^3} {k^\mu
k^\nu + 2\delta^{\mu\nu}p\cdot(p+k)\over p^2(p+k)^2}  }} After the
contraction with ${1\over 4}\epsilon_{\rho\mu\alpha} {k^\alpha\over
k^2} \epsilon_{\nu\sigma\beta} {k^\beta\over k^2}$, and dropping
linear divergences, \matterfipop\ becomes simply \eqn\emtt{ {1\over
32} { \delta_{\rho\sigma}k^2-k_\rho k_\sigma \over (k^2)^{3\over 2}
} } The full integral is now \eqn\sdgmee{\eqalign{& -{1\over
32}\sum_{i=0}^{n-1}\int {d^3k\over (2\pi)^3} {(k\cdot\Delta) k\cdot
(2p+k)-((2p+k)\cdot\Delta) k^2\over (k^2)^{3\over 2} (p+k)^2 }
((p+k)\cdot\Delta)^i (p\cdot\Delta)^{n-1-i} \cr & \longrightarrow
{1\over 32}\sum_{i=0}^{n-1}\int {d^3k\over (2\pi)^3}
{(2p+k)\cdot\Delta\over (k^2)^{1\over 2} (p+k)^2 }
((p+k)\cdot\Delta)^i (p\cdot\Delta)^{n-1-i} \cr &= {1\over
128\pi^2}(p\cdot\Delta)^n\ln\Lambda \sum_{i=0}^{n-1}\int_0^1 dx
(1+x) x^{i-{1\over 2}} \cr &\sim {1\over 128\pi^2} \ln(n)
(p\cdot\Delta)^n \ln\Lambda~~~~~~~(n\gg 1)  }}

\listrefs

\end

\appendix{A}{The 2-loop correction to $|\phi|^6$ coupling in abelian Chern-Simons-matter theory}

In section 2 we argued that the classical Lagrangian of ${\cal N}=2$
CS-matter theory is not renormalized, based on supersymmetry and the
non-renormalization property of the Chern-Simons coupling. It is
nevertheless instructive to see this in perturbation theory
diagrammatically. The standard perturbative non-renormalization
theorem forbids $|\phi|^2$ and $|\phi|^4$ terms from being
generated, since they could only come from a superpotential. The
cancellation of the beta functions for the $|\phi|^6$ coupling is
less trivial.

This is in contrast with another type of supersymmetric theories
with $|\phi|^6$ coupling, the Wess-Zumino model with $\Phi^4$
superpotential, which has a positive beta function and is IR free.
One might be surprised that in the CS-matter theory, as suitable
couplings to fermions and the Chern-Simons gauge field are included,
the beta function for $|\phi|^6$ coupling vanishes. This is a
consequence of {\sl both} supersymmetry and the non-renormalization
property of the Chern-Simons coupling.

Let us examine the beta function for the $|\phi|^6$ coupling in the
abelian theory, with $N_f$ matter fields, of charge
$q=\sqrt{4\pi/k}$. The $|\phi|^6$ coupling is $g=q^4/4$ at tree
level. The 1-loop correction vanishes trivially due to kinematics.
The leading possible nontrivial contributions come from 2-loop
corrections, of order $q^8\sim 1/k^4$. The 2-loop beta function for
$g$ is of the form \eqn\asbett{ \mu {dg\over d\mu} = {q^8\over
96\pi^2}b } Since we anticipate $b$ to be zero, we will assume tree
level values for the couplings on the RHS instead of expressing it
more generally as a function of $g$ and other couplings in the
theory (which are supposed to be related to $q$ by supersymmetry).
The relevant diagrams are summarized in figure 6 and 7. For
concision we did not exhibit the orientation and indices of the
graphs, which are of course important to get the combinatorical
factors. Let us make a few comments on the calculation.

\vskip 0.5cm \vbox{ \centerline{\vbox{ \hbox{\vbox{\offinterlineskip
\halign{&#&\strut\hskip0.2cm \hfill #\hfill\hskip0.2cm\cr
 ~~~~~~~~~~~& \epsfysize=0.6in \epsfbox{twoloop.1} && \epsfysize=0.6in \epsfbox{twoloop.2}
  && \epsfysize=0.6in \epsfbox{twoloop.3}  && \epsfysize=0.6in \epsfbox{twoloop.4} & \cr &
  $(a)$ && $(b)$ && $(c)$ && $(d)$ & \cr & & \cr &
  \epsfysize=0.6in \epsfbox{twoloop.5} && \epsfysize=0.6in \epsfbox{twoloop.6} && \epsfysize=0.6in \epsfbox{twoloop.7} &&
  \epsfysize=0.6in \epsfbox{twoloop.8} & \cr &
  $(e)$ && $(f)$ && $(g)$ && $(h)$ & \cr& &  \cr & \epsfysize=0.6in \epsfbox{twoloop.9} &&
  \epsfysize=0.6in \epsfbox{twoloop.10} && \epsfysize=0.6in \epsfbox{twoloop.11} && \epsfysize=0.6in \epsfbox{twoloop.12} &
  \cr &
  $(i)$ && $(j)$ && $(k)$ && $(l)$ &  \cr& &  \cr
  & \epsfysize=0.6in \epsfbox{twoloop.13} && \epsfysize=0.6in \epsfbox{twoloop.14} && \epsfysize=0.6in \epsfbox{twoloop.15} & && &\cr
  &
  $(m)$ && $(n)$ && $(o)$ && & \cr
}}} \centerline{{\bf Figure 6:} two loop corrections to the
$|\phi|^6$ coupling.}}} } \vskip 0.5cm

The Chern-Simons propagator in the Feynman gauge is \eqn\ascsprop{
{\epsilon_{\mu\nu\rho} k^\rho-i\xi {k_\mu k_\nu\over k^2}\over 2k^2}
} where $\xi$ is a gauge fixing parameter. We will work with the
choice $\xi=0$. The regularization of Chern-Simons perturbation
theory is a subtle issue \refs{\AlvarezGaumeWK,\ChenEE}. The most
natural regularization method is to introduce a Yang-Mills coupling
of the form ${1\over\Lambda} F_{\mu\nu}F^{\mu\nu}$. With this
regularization the Chern-Simons theory becomes super-renormalizable,
and there are only a finite number of diagrams that need to
regularized using other methods. The disadvantage is that the
propagator for the gauge field becomes rather complicated, and the
calculations are messy. An alternative consistent regularization
method is ``dimensional reduction" \ChenEE, where the tensor algebra
is performed in 3 dimensions, and the final integrals over momenta
can be performed in $3-\epsilon$ dimensions or with a suitable
cutoff. We will use dimensional reduction in this paper. This
regularization method leads to a different finite shift of the
Chern-Simons level from the YM regularization, but this difference
does not show up in the logarithmic divergence at two-loop.

We shall note a few simplifications in the diagrammatics. Firstly,
when the gauge field propagator is attached to an external scalar
propagator via a trivalent vertex, the diagram is zero due to the
contraction of the momentum at the vertex with the Chern-Simons
propagator \ascsprop. So the only way to attach gauge field
propagators to external scalars is through the quadrivalent vertex,
such as $(e)$ in figure 6. Secondly, straightforward calculation
reveals the cancelation between diagrams $(l)$ and $(m)$ in figure
6, as well as the cancelation of $(n),(o)$ in figure 6 against the
wave function renormalization due to $(d),(e)$ in figure 7.
Furthermore, the contributions from $(g)$ in figure 6 and $(f), (g)$
in figure 7 to the beta function vanish.

\vskip 0.5cm \centerline{\vbox{ \hbox{\vbox{\offinterlineskip
\halign{&#&\strut\hskip0.2cm \hfill #\hfill\hskip0.2cm\cr
 ~~~~~~~~~~~~~~~~~&  && \epsfysize=0.6in \epsfbox{wavefunction.1}
  &&   &&  & \cr & & \cr &
  \epsfysize=0.6in \epsfbox{wavefunction.2} && \epsfysize=0.6in \epsfbox{wavefunction.3} && \epsfysize=0.6in \epsfbox{wavefunction.4} &&
  \epsfysize=0.6in \epsfbox{wavefunction.5} & \cr &
  $(a)$ && $(b)$ && $(c)$ && $(d)$ & \cr & &  \cr & \epsfysize=0.6in \epsfbox{wavefunction.6} &&
  \epsfysize=0.6in \epsfbox{wavefunction.7} && \epsfysize=0.6in \epsfbox{wavefunction.8} &&  &
  \cr &
  $(e)$ && $(f)$ && $(g)$ &&  &  \cr
}}} \centerline{{\bf Figure 7:} the contribution due to wave
function renormalization.}}} \vskip 0.5cm

We can separate the calculation into two parts. First, we consider
the contribution of order $q^8 N_f$, in other words, the correction
to the $|\phi|^6$ coupling in the large $N_f$ limit with $q^4N_f\sim
N_f/k^2$ kept finite. Note that this is not the same as the standard
large $N_f$ 't Hooft-like limit, where $N_f/k$ is kept finite. In
the latter limit the 2-loop contribution to the beta function
vanishes trivially (and the theory can be solved by saddle point
approximation).

To order ${\cal O}(N_f)$ (i.e. the diagrams with one flavor index
summed over), the nonzero contributions to $b$ in \asbett\ from the
diagrams in figure 6 are (multiplied by $N_f$) \vskip 0.5cm \vbox{
\centerline{\vbox{ \hbox{\vbox{\offinterlineskip
\def\tablespace{height5pt&\omit&&\omit&&\omit&&\omit&&\omit&&\omit&&\omit&\cr}
\def\tablerule{\tablespace\noalign{\hrule}\tablespace}

\hrule\halign{&\vrule#&\strut\hskip0.2cm\hfill #\hfill\hskip0.2cm\cr
\tablespace & a && b  && c  && d && e && f && h &\cr
\tablerule &  ${27\over 2}$ && ${81\over 8}$  && $-{63\over 4}$  &&
$-{39\over 4}$ && $-{9\over 2}$ && $3$ && ${3\over 2}$ &\cr
\tablespace}\hrule}}}} } The nontrivial contribution to $b$ from
wave function renormalization, coming from only (a) in figure 7 at
order $N_f$, is calculated to be $15 N_f/8$. These add up to
precisely zero! Note that to this order, the diagrams involving only
the matter fields, namely $(a-d)$ in figure 1 together with the wave
function renormalization, cancel among themselves. It is curious
whether the matter sector of the theory could be exactly conformal
all by themselves in the limit $N_f\to\infty$, $N_f/k^2$ finite.

To determine the ${\cal O}(1)$ contribution, we can consider the
theory with $N_f=1$ (so that it is easy to keep track of the
combinatorical factors). The contributions to $b$ from the diagrams
in figure 6 are \vskip 0.5cm \vbox{ \centerline{\vbox{
\hbox{\vbox{\offinterlineskip
\def\tablespace{height5pt&\omit&&\omit&&\omit&&\omit&&\omit&&\omit&&\omit&&\omit&&\omit&&\omit&\cr}
\def\tablerule{\tablespace\noalign{\hrule}\tablespace}

\hrule\halign{&\vrule#&\strut\hskip0.2cm\hfill #\hfill\hskip0.2cm\cr
\tablespace & a && b  && c  && d && e && f && h && i && j && k &\cr
\tablerule &  $63$ && ${405\over 8}$  && $-{243\over 4}$  &&
$-{243\over 4}$ && $-{27\over 2}$ && $9$ && ${27\over 2}$ &&
$-{45\over 4}$ && $6$ && ${9\over 2}$ &\cr \tablespace}\hrule}}}} }
The contribution to $b$ from the wave function renormalization in
figure 7 are \vskip 0.5cm \vbox{ \centerline{\vbox{
\hbox{\vbox{\offinterlineskip
\def\tablespace{height5pt&\omit&&\omit&&\omit&\cr}
\def\tablerule{\tablespace\noalign{\hrule}\tablespace}

\hrule\halign{&\vrule#&\strut\hskip0.2cm\hfill #\hfill\hskip0.2cm\cr
\tablespace & a && b  && c  &\cr
\tablerule &  ${27\over 8}$ && $-{3\over 4}$  && $-3$  &\cr
\tablespace}\hrule}}}} } Once again, these add up to precisely zero!
Note that now the diagrams involving only matter fields do {\sl not}
cancel among themselves, but they cancel against the diagrams that
involve the Chern-Simons gauge field. In summary, we found that the
2-loop contribution to the beta function for the $|\phi|^6$ coupling
vanishes through explicit calculation, as previously anticipated.